\documentclass[twocolumn,times]{aastex62}
\usepackage{graphicx, amsmath, amssymb}
\usepackage{ulem}
\usepackage{color}

\received{December 12, 2017}
\revised{\today}
\accepted{\today}

\shorttitle{Statistical Signatures of Nanoflare Activity. I.}
\shortauthors{Jess et al.}

\begin{document}
\title{Statistical Signatures of Nanoflare Activity. I. \\ Monte Carlo Simulations and Parameter-space Exploration}
\correspondingauthor{D.~B. Jess}
\email{d.jess@qub.ac.uk}

\author[0000-0002-9155-8039]{D. B. Jess} 
\affiliation{Astrophysics Research Centre, School of Mathematics and Physics, Queen's University Belfast, Belfast, BT7 1NN, UK}                                  
\affiliation{Department of Physics and Astronomy, California State University Northridge, Northridge, CA 91330, USA}      
\author[0000-0003-2709-7693]{C. J. Dillon}
\affiliation{Astrophysics Research Centre, School of Mathematics and Physics, Queen's University Belfast, Belfast, BT7 1NN, UK}                                  
\author[0000-0001-9874-1429]{M. S. Kirk}
\affiliation{NASA Goddard Space Flight Center, Code 670, Greenbelt, MD 20771, USA}
\author[0000-0002-1820-4824]{F. Reale}
\affiliation{Dipartimento di Fisica \& Chimica, Universit{\`{a}} di Palermo, Piazza del Parlamento 1, I-90134 Palermo, Italy}
\affiliation{INAF-Osservatorio Astronomico di Palermo, Piazza del Parlamento 1, I-90134 Palermo, Italy}
\author[0000-0002-7725-6296]{M. Mathioudakis}
\affiliation{Astrophysics Research Centre, School of Mathematics and Physics, Queen's University Belfast, Belfast, BT7 1NN, UK}
\author[0000-0001-5170-9747]{S. D. T. Grant} 
\affiliation{Astrophysics Research Centre, School of Mathematics and Physics, Queen's University Belfast, Belfast, BT7 1NN, UK}     
\author[0000-0003-1746-3020]{D. J. Christian}
\affiliation{Department of Physics and Astronomy, California State University Northridge, Northridge, CA 91330, USA}
\author[0000-0001-8556-470X]{P. H. Keys}
\affiliation{Astrophysics Research Centre, School of Mathematics and Physics, Queen's University Belfast, Belfast, BT7 1NN, UK}
\author[0000-0002-0735-4501]{S. Krishna Prasad}
\affiliation{Astrophysics Research Centre, School of Mathematics and Physics, Queen's University Belfast, Belfast, BT7 1NN, UK}
\author[0000-0001-5547-4893]{S. J. Houston} 
\affiliation{Astrophysics Research Centre, School of Mathematics and Physics, Queen's University Belfast, Belfast, BT7 1NN, UK}

\begin{abstract}
Small-scale magnetic reconnection processes, in the form of nanoflares, have become increasingly hypothesized as important mechanisms for the heating of the solar atmosphere, for driving propagating disturbances along magnetic field lines in the Sun's corona, and for instigating rapid jet-like bursts in the chromosphere. Unfortunately, the relatively weak signatures associated with nanoflares places them below the sensitivities of current observational instrumentation. Here, we employ Monte Carlo techniques to synthesize realistic nanoflare intensity time series from a dense grid of power-law indices and decay timescales. Employing statistical techniques, which examine the modeled intensity fluctuations with more than 10$^{7}$ discrete measurements, we show how it is possible to extract and quantify nanoflare characteristics throughout the solar atmosphere, even in the presence of significant photon noise. A comparison between the statistical parameters (derived through examination of the associated intensity fluctuation histograms) extracted from the Monte Carlo simulations and SDO/AIA 171{\,}{\AA} and 94{\,}{\AA} observations of active region NOAA~11366 reveals evidence for a flaring power-law index within the range of $1.82 \leq \alpha \leq 1.90$, combined with $e$-folding timescales of $385\pm26${\,}s and $262\pm17${\,}s for the SDO/AIA 171{\,}{\AA} and 94{\,}{\AA} channels, respectively. These results suggest that nanoflare activity is not the dominant heating source for the active region under investigation. This opens the door for future dedicated observational campaigns to not only unequivocally search for the presence of small-scale reconnection in solar and stellar environments, but also quantify key characteristics related to such nanoflare activity.
\end{abstract}


\section{Introduction}
Magnetic reconnection is a common process within solar and stellar atmospheres. During reconnection phenomena, magnetic fields are rearranged into a stable state of lower energy, thus releasing a considerable excess in the form of increased kinetic energies of the embedded plasma, the acceleration of charged particles and extreme localized heating \citep{Pri86, Pri99, Hudson1991, Hud11}. It is the deposition of thermal energy that has been postulated as one of the main mechanisms for supplying the background heat flux necessary to maintain the multi-million degree temperatures present in the outer solar atmosphere. The signatures of such events can readily be observed during the impulsive stages of large-scale flares, which can often release in excess of $10^{31}$~erg of energy within a compact volume. However, the relative rarity of large flares, particularly during periods of solar minima, means that they cannot solely provide the sustained heating required. As a result, nanoflares were proposed whereby smaller (individual energies on the order of $10^{24}$~erg), yet more frequent magnetic reconnection events may be able to remain active throughout the extremities of the solar cycle, while also providing a continued basal background heating \citep{Parker88}. 

In order for such a mechanism to be dominant, the occurrence rate of nanoflares must be substantially higher than those for larger-scale flaring events. The continuous spread of flaring energies are believed to be governed by a power-law relationship, whereby the frequency, $dN/dE$, of flaring events with an associated energy, $E$, is described by,
\begin{equation*}
\frac{dN}{dE} \sim E^{-\alpha} \ ,
\end{equation*} 
where $\alpha$ is the power-law index. It is required that $\alpha${\,}$\ge${\,}$2$ for nanoflares to play an important role in the heating of the solar atmosphere \citep{Parker88, Hudson1991}. Unfortunately, while measurements of the power-law index for large-scale flares are relatively straightforward, observational constraints can often introduce significant errors in the calculation of a power-law index applicable to lower energy events. Such constraints have been documented by \citet{Hannah08}, who suggest that frequency turnovers at low energies may be caused by instrumental effects as a consequences of missing (or failing to detect) the smallest events. As a result, often the largest uncertainties in the derived power-law indices are associated with nanoflare type events, with estimations spanning $1.35${\,}$\le${\,}$\alpha${\,}$\le${\,}$2.90$ \citep{Berg98, Krucker98, Aschwanden1999, Parnell00, Benz02, Wine02, Aschwanden2012, Aschwanden2014, Aschwanden2015}. 

Furthermore, \citet{Lopez07} have demonstrated that the intensity of an impulsively heated coronal loop must be a direct indication of the nanoflare occurrence rate, whereby small-scale energies could be injected frequently, or larger energies may be introduced more intermittently, thus opening up the possibility that individual structures may be governed by either a traditional range of flare energies and occurrence rates (i.e., following a power law), or by a narrow range of energies being applied more regularly in time. As a result, it is presently unclear whether nanoflare energies and occurrences are significant enough to be a dominant contributor to atmospheric heating. Nevertheless, in more recent years, nanoflares have also been proposed as viable mechanisms to initiate magneto-hydrodynamic wave activity in the chromosphere \citep{Kli14} and corona \citep{Wang13}, while more extreme examples of nanoflare activity may be responsible for heating chromospheric plasma to transition region temperatures in the form of Type~{\sc{ii}} spicules \citep{Bra15}.

As highlighted above, the energetics associated with nanoflares places them on, or below, the sensitivity limits of current telescope facilities and instrumentation. As a result, ongoing research is attempting to devise novel ways to diagnose, extract and characterize nanoflares from data that often display no clear impulsive signatures. Current approaches include the use of spectroscopic techniques to compare the scaling between kinetic temperatures and emission measures of coronal plasma \citep[{\rmfamily e.g.,}][]{Kli01, Bra12}. \citet{Sar08, Sar09} employed a multi-stranded loop model and folded their synthetic outputs through EUV instrumental response functions to examine whether the resulting emission-measure-weighted temperature profiles could be conclusively examined for the presence of nanoflare activity. The authors found that broad differential emission measures were produced, but that any potential observational signatures may be below the detection thresholds of current EUV imaging instrumentation. Such limitations may result from what is termed the ``isothermal bias'', where \citet{Web05} utilized a flat differential emission measure distribution to mimic an inherently multi-thermal plasma and revealed that filter ratio methods used to construct the differential emission measures are biased towards the temperature response functions of the imaging channels used. Then, as a result of the electron temperatures and the thermal energies being statistically correlated during flare processes, \citet{Aschwanden2002} also demonstrated how emission measure approaches spanning a limited temperature range naturally introduce a bias in the frequency distribution of flare energies, thus affecting the derived power-law index. As a consequence, the reliability of such approaches hinge upon the accurate diagnosis of isothermal and multi-thermal plasma when constructing the emission measures, as well as the number of optically-thin magnetic strands superimposed along the observational line-of-sight. Indeed, \citet{Car14} recently demonstrated how the flare energy power law derived from differential emission measure techniques is sensitive to the time between individual nanoflares, suggesting that the associated energies may be smaller than previously envisioned. Furthermore, \citet{Reale08} employed simulations to document how non-equilibrium ionization effects during the heating stages of nanoflare activity may result in the undetectability of heat pulses shorter than approximately 1~minute in duration, which naturally affects the ability of differential emission measure techniques to extract the signatures of short-lived nanoflare events. Nevertheless, \citet{Reep13} utilized a hydrodynamic model to reveal how steady heating (i.e., where the timescale between events is shorter than the cooling timescale) may be able to replicate 86\% to 100\% of active region core emission measures where nanoflare heating may be prevalent. \citet{Reep13} also add a caveat that the slopes of the emission measures deduced from observations alone are not sufficient to provide information about the specific timescales associated with heating. However, more recently, \citet{Ishikawa2017} employed differential emission measure techniques on hard X-ray observations from the second flight of the Focusing Optics X-ray Solar Imager \citep[FOXSI--2;][]{Krucker2009, Christe2016} sounding rocket, and revealed plasma heated above 10~MK, thus providing yet more evidence for the existence of solar nanoflares.

\citet{Sak08} and \citet{Vek09} compared co-temporal intensity time series obtained at EUV and X-ray wavelengths, and estimated that a `hot' corona could be maintained with nanoflare filling factors on the order of 10\%. Subsequent theoretical modeling by \citet{Joshi12} provided corroborating evidence that the X-ray fluctuations observed by \citet{Kat01} and \citet{Sak08} could be representative of 10$^{23}$ -- 10$^{26}$~erg events released over timescales of $\sim$100{\,}s. Importantly, the results of \citet{Sak08} demonstrate a lag time between soft X-ray and EUV time series (corresponding to the cooling timescale), which suggests that soft X-ray loops may require higher nanoflare energies than their EUV counterparts, thus perhaps indicating a wavelength dependence on the nanoflare power-law index. However, the observations employed by \citet{Sak08} and \citet{Vek09} were previous generation TRACE \citep{Han99} and Yohkoh/SXT \citep{Oga91, Tsu91} images with reduced (by modern standards) spatial and temporal resolutions, which as a result, limited the statistical significance of their results. Employing modern EUV image sequences acquired by the Atmospheric Imaging Assembly \citep[AIA;][]{Lem12} onboard the Solar Dynamics Observatory \citep[SDO;][]{Pes12}, a series of papers by \citet{Viall11, Viall12, Viall13, Viall15, Viall16, Viall17} examined the intensity fluctuations captured in the high-resolution EUV time series. \citet{Viall11} documented fluctuations in the associated lightcurves on times scales of $\sim$20 minutes, and concluded that this was inconsistent with a steady heating model as a result of the clearly impulsive nature of the extracted time series. Furthermore, time delays between the impulsive signatures found by \citet{Viall12} in the temperature sensitive EUV observations corroborated the cooling plasma interpretation of \citet{Sak08}, and also further suggested the presence of impulsive nanoflare heating.

\citet{Ter11} and \citet{Jess14} offered an alternative approach to identify nanoflare signatures embedded within long-duration solar time series. These authors utilized direct imaging techniques to build up a statistical picture of small-scale fluctuations contained within the pixel lightcurves. Through comparison with Monte Carlo simulations, subtle asymmetries of the measured intensity fluctuations could be interpreted as the signatures of successive impulsive events embedded within an inherently cooling plasma. Since large number statistics were employed in the studies by \citet{Ter11} and \citet{Jess14}, instrumental effects due to calibration uncertainties are likely to be minimized due to their very small fluctuations around the relevant mean. Hence, such instrumental effects are very unlikely to introduce large-scale intensity fluctuations that produce the broad (and often asymmetric) tails of the corresponding intensity fluctuation histograms. As a result, \citet{Ter11} and \citet{Jess14} both linked such statistical signatures to the presence of real nanoflare events, something that was previously put forward by \citet{Kat01}.

Importantly, this technique does not rely on the presence of optically thin observations (i.e., coronal observations), nor does it require the accurate fitting of multi-thermal plasma properties, thus also making it suitable for studies of the lower solar atmosphere. In a series of publications examining whether such small-scale flare events could heat stellar coronae, \citet{Gudel97}, \citet{Kas02}, \citet{Gudel03} and \citet{Arzner04} compared a similar time-dependent Poisson process of impulsive events to Extreme Ultraviolet Explorer/Deep Survey \citep[EUVE/DS;][]{Malina91} observations of the flaring star AD~Leo. Time series of AD~Leo has shown continuous variability \citep[e.g.,][]{Ambruster87}, which has been suggested as the observational signature of a large number of superimposed flares characterized by similar decay timescales. The simulated EUV and X-ray time series of \citet{Arzner04} revealed that the expected count rates (extracted from the Fourier transform of the flare probability densities) were related to the mean flaring interval, suggesting the occurrence of small-scale events may have a (quasi-)periodic dependency. Furthermore, the comparison of such synthetic time series to the EUVE/DS observations indicated a flaring power-law index of $\sim$2.3, suggesting nanoflares may play an important role in the heating of both solar and stellar coronae.

\begin{figure*}
\begin{center}
\includegraphics[width=0.6\textwidth, clip=true]{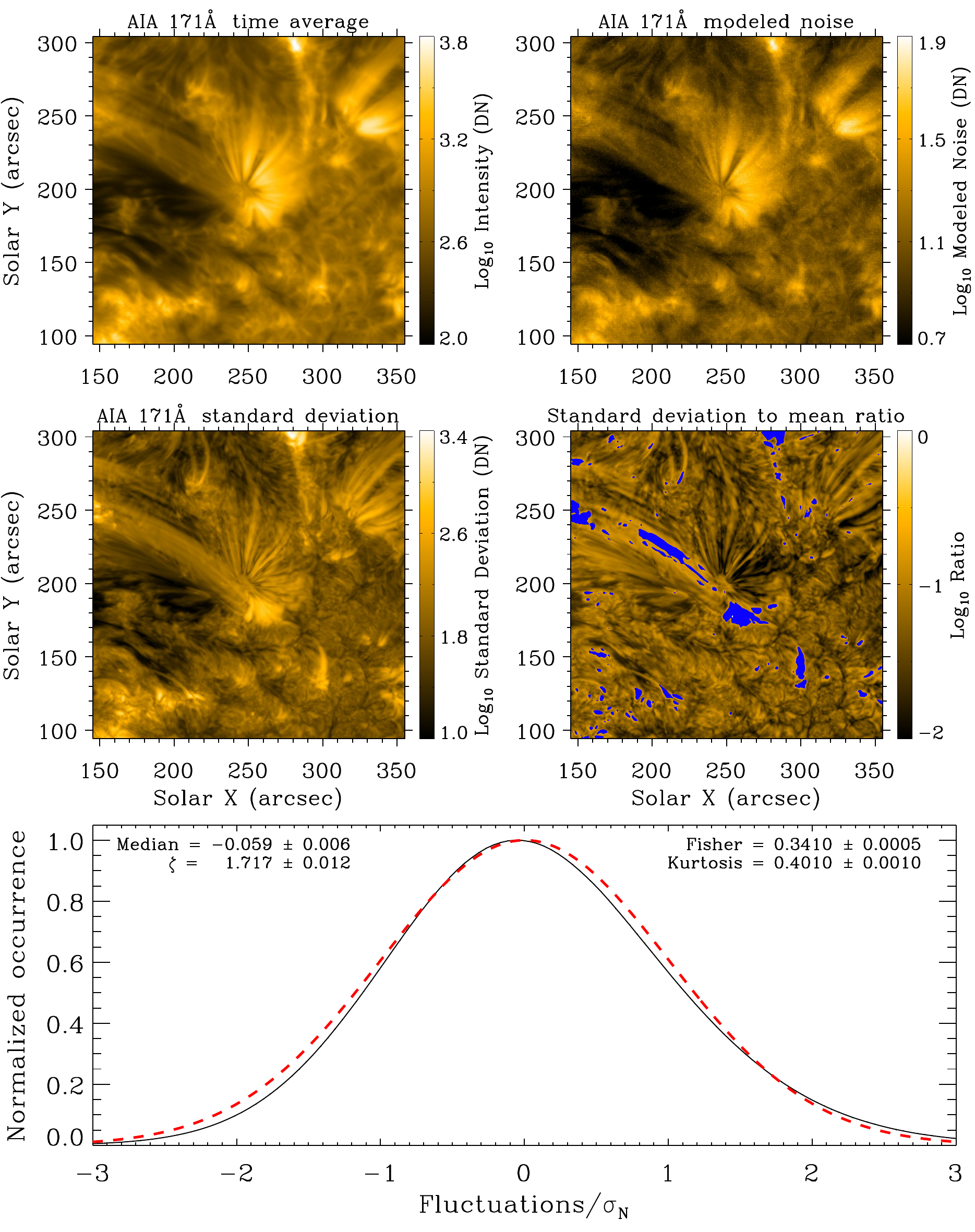}
\end{center}
\caption{A $210{\arcsec}\times210{\arcsec}$ (350{\,}$\times${\,}350 pixels$^{2}$) SDO/AIA 171{\,}{\AA} snapshot averaged in time across 15:30 -- 18:00~UT on 2011 December 10 (upper-left panel), which is used as the base science image for the Monte Carlo simulations. The upper-right panel displays the magnitude of Gaussian--Poisson noise fluctuations estimated from the 171{\,}{\AA} time series. The middle-left image depicts the standard deviations of the SDO/AIA 171{\,}{\AA} intensity time series, while the middle-right image reveals the $\frac{\mathrm{standard~deviation}}{\mathrm{mean}}$ ratios across the same field-of-view. Regions demonstrating a ratio larger than 0.20 are removed from the present study, with those pixels masked out using blue contours. All images are displayed on the solar heliocentric co-ordinate system using a log-scale to better reveal fine-scale structuring that would otherwise be swamped by the large intensity ranges between the brightest and darkest features. The lower panel displays the corresponding intensity fluctuation distribution (black line; in units of $\sigma_{N}$), normalized to its maximum, for the SDO/AIA~171{\,}{\AA} field-of-view. A standardized Gaussian profile is also plotted as a red dashed line for reference.
}
\label{SDO171_images}
\end{figure*}

Unfortunately, one drawback of statistical approaches is the fact that the inferred nanoflare characteristics only become most accurate in the limit of large number statistics. Since the solar atmosphere is constantly evolving through oscillatory phenomena, structural dissipation, feature drifts alongside new magnetic flux emergence, such approaches naturally require a combination of a well-defined region of interest, in addition to large number statistics to help remove the contributions of non-nanoflare phenomena from the resulting intensity distributions. Thankfully, high-cadence and long-duration imaging sequences are now commonplace, particularly from ground-based observatories that are not limited by the same telemetry restrictions as space-borne instrumentation. Specifically, the work of \citet{Ter11} utilized approximately $2\times10^{7}$ individual measurements by employing the X-Ray Telescope \citep[XRT;][]{Gol07} onboard Hinode, while \citet{Jess14} increased this limit to over $1\times10^{9}$ discrete values by employing the Hydrogen-Alpha Rapid Dynamics camera \citep[HARDcam;][]{Jess12} on the Dunn Solar Telescope. More recently, \citet{Taj16a, Taj16b} utilised EUV image sequences captured by SDO/AIA to examine small-scale intensity fluctuations using probabilistic neural networks and cross-correlation techniques. \citet{Taj16a, Taj16b} conclude that maps of pixel intensity fluctuations, as previously demonstrated by \citet{Ter11} and \citet{Jess14}, may provide excellent diagnostic capabilities for deducing nanoflare characteristics in the solar atmosphere.

Even with such large measurement numbers, several unresolved statistical features manifested as a result of the analyses of \citet{Ter11} and \citet{Jess14}. While both intensity fluctuation histograms were negatively offset from the mean, they also displayed a degree of positive skewness in their composition. Furthermore, the widths of the distributions were not completely aligned with a standardized Gaussian profile, something which would be expected for time series entirely comprised of photon noise statistics. Therefore, a significant number of questions arose, and remain unanswered, as a result of the above mentioned work. While many of these issues have not been directly addressed, \citet{Jess14} suggested that larger impulsive events \citep[e.g., perhaps related to the lower-energy microflares documented by][]{Jes10a} may result in more contributions to larger intensity fluctuations causing them to remain elevated over a wider range of values, thus inducing a degree of positive skewness in the statistical distributions. Here, we take the hypotheses put forward by \citet{Jess14} one step further by analyzing a series of Monte Carlo simulations that are designed to replicate the intensity perturbations (both impulsive and decay signatures) caused by small-scale nanoflare activity in the solar corona. Such synthesized time series are constructed using a dense grid of input parameters, which includes the underlying power-law index and the associated decay timescales, to ascertain the plasma and nanoflare characteristics responsible for pronounced asymmetries in the resulting statistical distributions. In addition, we utilize high-resolution multi-wavelength SDO/AIA observations of a quiescent active region to investigate how the modeled statistical parameters and synthetic time series compare to their observational counterparts.

\begin{figure*}
\begin{center}
\includegraphics[width=0.6\textwidth, clip=true]{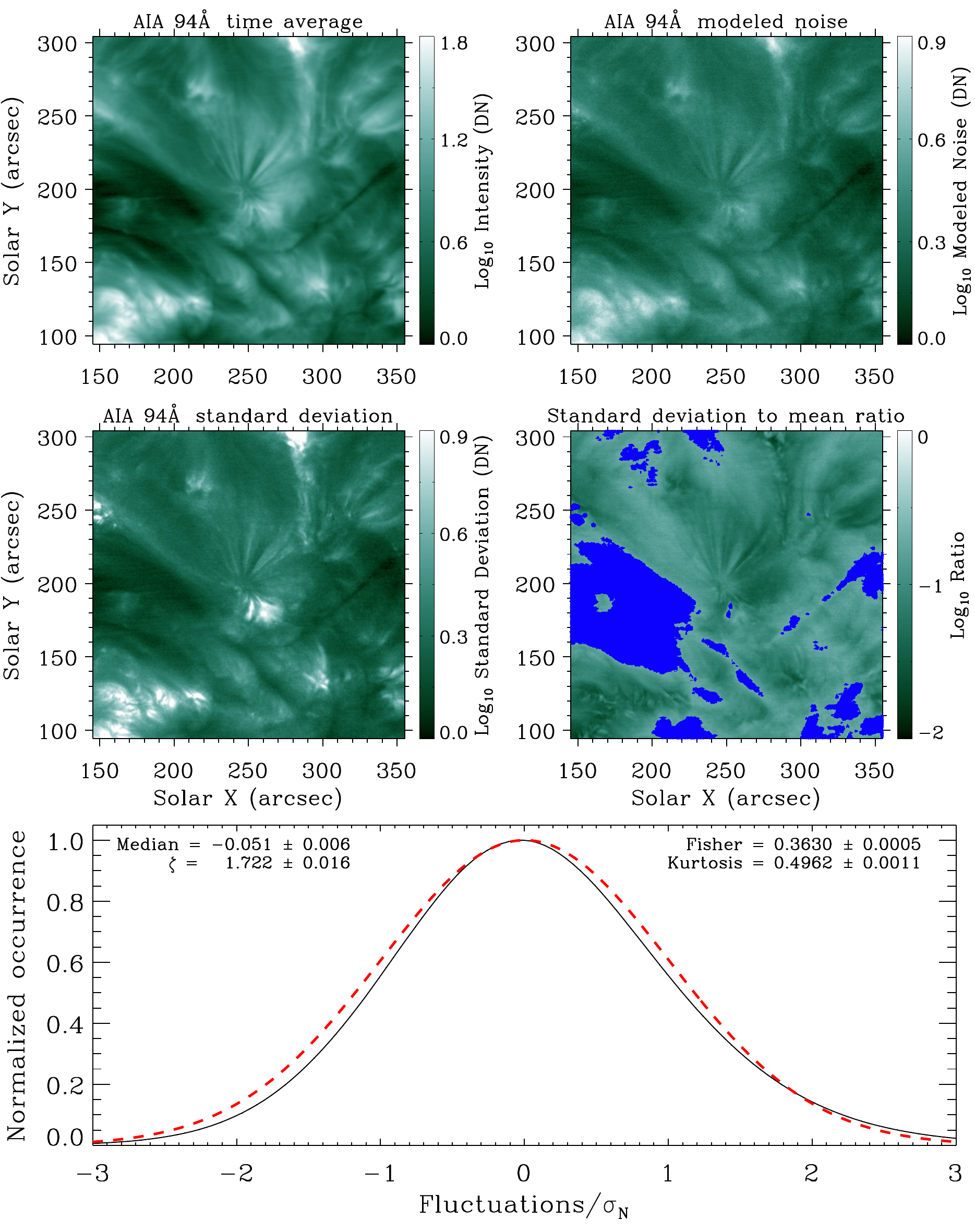}
\end{center}
\caption{In an identical way to Figure~{\ref{SDO171_images}}, the upper-left panel displays a time-averaged SDO/AIA 94{\,}{\AA} image, while the upper-right panel depicts the magnitude of Gaussian--Poisson noise fluctuations estimated from the 94{\,}{\AA} time series. The middle-left image highlights the standard deviations of the SDO/AIA 94{\,}{\AA} intensity time series, while the middle-right image reveals the $\frac{\mathrm{standard~deviation}}{\mathrm{mean}}$ ratios across the same field-of-view. Regions demonstrating a ratio larger than 0.35 are removed from the present study, with those pixels masked out using blue contours. The lower panel displays the corresponding intensity fluctuation distribution (black line; in units of $\sigma_{N}$), normalized to its maximum, for the SDO/AIA~94{\,}{\AA} field-of-view. A standardized Gaussian profile is also plotted as a red dashed line for reference.}
\label{SDO94_images}
\end{figure*}

\section{Observations}
\label{observations}
The active region employed throughout this study is NOAA~11366, with the time sequence spanning 15:30 -- 18:00~UT on 2011 December 10, and comprising of 750 171{\,}{\AA} images, alongside 750 counterpart 94{\,}{\AA} images, each with a cadence of 12{\,}s. A $210{\arcsec}\times210{\arcsec}$ (350{\,}$\times${\,}350 pixels$^{2}$) field-of-view is employed to encapsulate the majority of the active region core. These data products have been extensively documented in previous published work, with further details available in \citet{Krishna15, Krishna16} and \citet{Jess16}. A time-averaged 171{\,}{\AA} image, visible in the upper-left panel of Figure~{\ref{SDO171_images}}, depicts a collection of loop and fan structures extending outwards from the underlying sunspot, with the surrounding areas displaying quiescent moss and background coronal plasma. Similar coronal features, albeit with reduced signal, are visible in the time-averaged 94{\,}{\AA} image displayed in the upper-left panel of Figure~{\ref{SDO94_images}}. The SDO/AIA 171{\,}{\AA} and 94{\,}{\AA} channels were chosen for this preliminary study due to their dominant sensitivities to coronal plasma below and above, respectively, a temperature of 1~MK \citep{Boerner2012}. During the 2.5~hour observations, no large-scale (macroscopic) flare brightenings were observed in the SDO/AIA data, highlighting the quiescent nature of the active region under investigation. 

The particular active region was chosen since it was a decaying (McIntosh classification Cao; Hale class $\beta\gamma\delta$) sunspot group that did not provide any large-scale (i.e., {\it{GOES}} C class or above) activity both during, and immediately prior to the observations being acquired. Studies undertaken by \citet{1990SoPh..125..251M} and \citet{2016ApJ...820L..11J} estimate the occurrence of McIntosh Cao and Hale $\beta\gamma\delta$ sunspot classifications as $\sim$7\% and $\sim$5\%, respectively. Therefore, they are not rare sunspot groups, and indeed form regularly throughout the solar cycle, but with a slight preference for the stages leading up to solar maxima \citep{2016ApJ...820L..11J}. The last detected activity from NOAA~11366 was when this active region produced a small C1.9 class flare on 2011~December~05 (when the sunspot group was McIntosh class Hsx, or Hale class $\alpha\gamma\delta$), some 5 days prior to our observational dataset. The relative quietness of the active region was deemed important when attempting to undertake a statistical study of nanoflare activity. Large, macroscopic flaring events would either need to be removed from the time series (reducing the useable time intervals and therefore affecting the statistical distributions), or masked out from the subsequent datacube (decreasing the number of available pixels for analysis). Applying such processes would naturally reduce the statistical significance of the derived interpretations. As such, we opted for a quiescent active region, in the form of NOAA~11366, that demonstrated no macroscopic flaring events.

\begin{figure*}
\begin{center}
\includegraphics[width=0.5\textwidth, trim={2cm 2cm 2cm 0cm}, angle=270, clip=true]{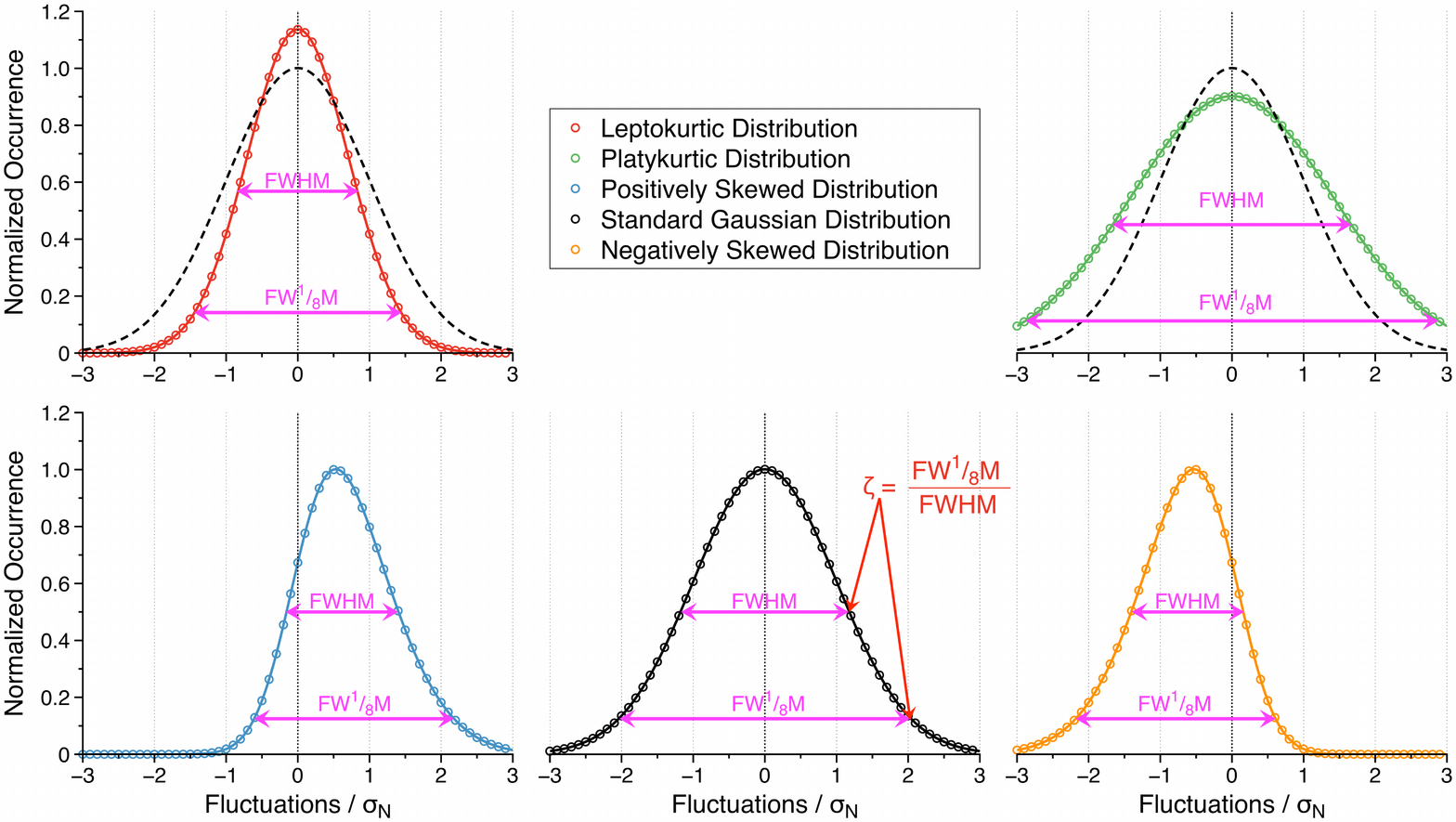}
\end{center}
\caption{
A collection of synthetic distributions detailing the shapes associated with leptokurtic (red; upper-left), platykurtic (green; upper-right), positively skewed (blue; lower-left) and negatively skewed (orange; lower-right) histograms. Each distribution is plotted as a function of the standard deviation, but is here labeled as $\sigma_{N}$ to remain consistent with the nomenclature used in the lower panels of Figures~{\ref{SDO171_images}}~\&~{\ref{SDO94_images}}. The leptokurtic and platykurtic distributions are plotted relative to a standardized Gaussian (dashed black line) to better show the changes to occurrences these distributions induce, while the skewed and Gaussian distributions (lower panels) are each normalized by their own maximum occurrence for clarity. The pink arrows highlight the full widths associated with each distribution at half- and eighth-maxima, while the red arrows reveal how the $\zeta$ (FW$\frac{1}{8}$M-to-FWHM ratio) term is calculated, which equals $1.73$ in the case of a standardized Gaussian.
}
\label{histogram_pics}
\end{figure*}

In order to better quantify the quiescence of the SDO/AIA active region core time series, we followed the methodology put forward by \citet{Lopez2016}. Here, the authors calculated ratios between the lightcurve standard deviations and means corresponding to SDO/AIA 171{\,}{\AA} time series, and through comparisons with Enthalpy-based Thermal Evolution of Loops \citep[EBTEL;][]{Klimchuk2008, Cargill2012} models, demonstrated how a ratio within the range of 0.08{\,}--{\,}0.20 is typical of SDO/AIA 171{\,}{\AA} image sequences displaying no large-scale flaring or obvious long-term variations. Therefore, an upper limit for the standard-deviation-to-mean ratio of 0.20 is placed upon our 171{\,}{\AA} time series, where regions displaying a ratio larger than 0.20 are masked out using blue contours in the middle-right panel of Figure~{\ref{SDO171_images}}. In total, 4309 pixels ($\sim$3\% of the field-of-view) are excluded from subsequent analysis, leaving the remaining 118{\,}191 spatial pixels for subsequent statistical investigation. Due to the much weaker signal-to-noise associated with the SDO/AIA 94{\,}{\AA} channel, a standard-deviation-to-mean ratio of 0.20 resulted in too much of the field-of-view being excluded. As a result, a standard-deviation-to-mean ratio of 0.35 was adopted for the SDO/AIA 94{\,}{\AA} channel to allow more pixels within the vicinity of the active region to be incorporated, while still remaining close to the threshold values put forward by \citet{Lopez2016}. This value resulted in 17{\,}460 pixels ($\sim$14\% of the field-of-view) being excluded from subsequent SDO/AIA 94{\,}{\AA} analysis, which can be identified by the blue contours in the middle-right panel of Figure~{\ref{SDO94_images}}.

Following the extraction of the SDO/AIA 171{\,}{\AA} and 94{\,}{\AA} pixels that conformed to the upper standard-deviation-to-mean ratios put forth by \citet{Lopez2016}, the time-resolved intensity fluctuations, $dI$, are computed similarly to \citet{Ter11} and \citet{Jess14},
\begin{equation}
dI(t) = \frac{I(t) - I_{0}(t)}{\sigma_{N}} \ ,
\label{eqn:inten_normalization}
\end{equation}
where $I(t)$ and $I_{0}(t)$ are the registered count and value of a linear least-squares fit used to de-trend the data, respectively, at time $t$, and $\sigma_{N}$ is an estimate of the time series noise, which in the limit of Poisson statistics is approximately equal to the standard deviation of the normalized pixel lightcurve. This is consistent with the natural shot noise distribution that arises from the particle nature of incident photons \citep[i.e.,][]{Ter77}, which is based around Poisson statistics \citep[for an in-depth overview see, e.g.,][]{Del08}. Note that the normalization performed above is similar to the Z-scores statistical transformation, which has widespread applications in physical and social sciences \citep{Sprinthall2012}.

A histogram of all $dI$ values is then computed, which by definition has a statistical mean equal to zero (see, e.g., the bottom panels of Figures~{\ref{SDO171_images}} \&~{\ref{SDO94_images}}). In order to best characterize the resulting distributions, a number of statistical parameters are evaluated, including the median offset, Fisher and Pearson coefficients of skewness, a measurement of the kurtosis and the variance of the histogram, along with the width of the distribution at a variety of locations, including at half-maximum, quarter-maximum, eighth-maximum, etc. Such statistical parameters have been employed in a variety of astrophysical research, including those linked to cosmological density fields \citep[e.g.,][]{1998A&A...331..829G}, quantifying polarization signals in radio observations \citep[e.g.,][]{2018MNRAS.474.3280F}, examining planetary orbits \citep[e.g.,][]{2006PASP..118.1510F} and uncovering the temporal evolution of large-scale solar flares \citep[e.g.,][]{2014SoPh..289..193S}. The ratio between the width of the distribution at eighth-maximum to that at half-maximum (i.e., FW$\frac{1}{8}$M-to-FWHM ratio) is defined here as `$\zeta$' for simplicity. A standardized Gaussian distribution has a natural ratio of $\zeta=1.73$. It must be stressed that the $\zeta$ measurement is distinctly different to the value of kurtosis, since kurtosis is a descriptor of the shape of the entire intensity fluctuation distribution, while the $\zeta$ parameter defines the relative widths of the distribution at distinct locations, in this case at the FWHM and the FW$\frac{1}{8}$M. Figure~{\ref{histogram_pics}} provides a graphical overview of the key statistical measurements employed here, alongside a standardized Gaussian distribution for comparison.  

For the SDO/AIA 171{\,}{\AA} dataset under current investigation, it is clear from the lower panel of Figure~{\ref{SDO171_images}} that a negative median offset ($-0.059\pm0.006$) is combined with a distribution that is narrower than that of a standardized Gaussian (i.e., $\zeta<1.73$, or more precisely $\zeta = 1.717\pm0.012$). Furthermore, the intensity fluctuation distribution is also leptokurtic (providing a positive value for the kurtosis; $0.4010\pm0.0010$) and slightly positively skewed (Fisher value of $0.3410\pm0.0005$). The SDO/AIA 94{\,}{\AA} dataset displays similar overall characteristics (lower panel of Figure~{\ref{SDO94_images}}), but with a reduced median offset ($-0.051\pm0.006$) and a slightly larger standardized width ($\zeta = 1.722\pm0.016$). When compared to that from the SDO/AIA 171{\,}{\AA} channel, the SDO/AIA 94{\,}{\AA} intensity fluctuation distribution is more leptokurtic with a kurtosis value of $0.4962\pm0.0011$, as well as more heavily skewed with a Fisher value equal to $0.3630\pm0.0005$. 

It must be noted that the median offset values have a high degree of precision, which is a consequence of both the large overall number statistics (approaching $10^{8}$ individual measurements) and the leptokurtic distributions placing the vast majority of fluctuations close to the natural mean of zero. To remain consistent with proven statistical methods \citep[e.g.,][]{Ken98, Tab06}, we adopt the standard skewness errors as $\sqrt{6/n}$, while the standard kurtosis errors are given by $\sqrt{24/n}$, where $n$ is the sample size used in the calculations. Thus, an individual SDO/AIA lightcurve, consisting of 750 discrete data points, provides Fisher skewness and kurtosis errors equal to $\pm0.175$ and $\pm0.351$, respectively, adopting a 95\% confidence interval. When the Fisher skewness and kurtosis errors are combined with the included 118{\,}191 SDO/AIA 171{\,}{\AA} pixels and 105{\,}040 SDO/AIA 94{\,}{\AA} pixels, these errors drop to $\pm0.0005$ and $\pm0.0010$ (171{\,}{\AA}) and $\pm0.0005$ and $\pm0.0011$ (94{\,}{\AA}), respectively. As one would expect from the work of \citet{Ken98} and \citet{Tab06}, the standard errors associated with the measurement of kurtosis are larger than those related to the Fisher skewness. The distributions depicted in the lower panels of Figures~{\ref{SDO171_images}} \&~{\ref{SDO94_images}} clearly deviate from those of standardized Gaussians, in terms of width, position and shape, and thus provide ideal testbeds to compare with the statistical fluctuations intrinsic to Monte Carlo nanoflare simulations.

\begin{table*}
\centering
\caption{Input parameters selected for the Monte Carlo simulation runs discussed in Section~{\ref{power-law_start}}.}
\label{tab:table1}
\begin{tabular}{@{}lcc@{}}
\hline
{\bf{Input}}		& {\bf{SDO/AIA}}  			& {\bf{SDO/AIA}}		\\
{\bf{Variable}}	& {\bf{171{\,}{\AA}}}			& {\bf{94{\,}{\AA}}}	\\
\hline
Background intensity (DN)	& Time-averaged SDO/AIA 171{\,}{\AA} image				& Time-averaged SDO/AIA 94{\,}{\AA} image	 \\
									& \footnotesize{(upper-left panel of Figure~{\ref{SDO171_images}})} 	& \footnotesize{(upper-left panel of Figure~{\ref{SDO94_images}})} \\
\vspace{-3mm} & & \\
Standard deviation of the noise ($\sigma_{N}$)	& Gaussian--Poisson noise estimation$^{a,b}$	& Gaussian--Poisson noise estimation$^{a,b}$	 \\
													& \footnotesize{(upper-right panel of Figure~{\ref{SDO171_images}})}		& \footnotesize{(upper-right panel of Figure~{\ref{SDO94_images}})} \\
\vspace{-3mm} & & \\
Dimensions of the simulated dataset (pixels)~~~~~~~~~~~~		& [$350, 350, 750$]						& [$350, 350, 750$]	 \\
 & \footnotesize{(original SDO/AIA 171{\,}{\AA} datacube size)} & \footnotesize{(original SDO/AIA 94{\,}{\AA} datacube size)} \\
\vspace{-3mm} & & \\
Flare energy power-law indices ($\alpha$)				& $1.5 \leq \alpha \leq 2.5$		& $1.5 \leq \alpha \leq 2.5$	 \\
																& \footnotesize{(steps of $0.02$)} & \footnotesize{(steps of $0.02$)} \\
\vspace{-3mm} & & \\
Nanoflare $e$-folding times (s)				& $10 \leq \tau \leq 1000$					& $10 \leq \tau \leq 1000$	 \\
	&  \footnotesize{(steps of $10$)} & \footnotesize{(steps of $10$)} \\
\vspace{-3mm} & & \\
Minimum simulated energy	(erg)					& $10^{22}$						& $10^{22}$	 \\
\vspace{-3mm} & & \\
Maximum simulated energy (erg)					& $10^{25}$						& $10^{25}$ \\
\vspace{-3mm} & & \\
1{\,}$\sigma_{N}$ flare energy (erg)				& $5\times10^{24}$						& $5\times10^{24}$ \\
\vspace{-3mm} & & \\
Surface area per pixel	(cm$^{2}$)	& $1.89225\times10^{15}$	& $1.89225\times10^{15}$ \\
\hline
\footnotesize $^{a}$: \citet{Kirk14a} & & ~~~~~~~~~~~~~~~~~~~~~~~~~~~~~~~~~~~~~~~~~~~~~~~~~~~~~~~~~~~~~~~~~~~~~~~~~ \\
\footnotesize $^{b}$: \citet{Kirk14b} & & \\
\end{tabular}
\end{table*}

\section{Monte Carlo Simulations}
\label{power-law_start}
In this Section we present a time series synthesis code that employs a traditional power-law distribution to govern the energetics and occurrences of time series signatures associated with nanoflares\footnote{A copy of the code can be obtained directly from D.B. Jess (\href{mailto:d.jess@qub.ac.uk}{d.jess@qub.ac.uk}), or by visiting \href{http://star.pst.qub.ac.uk/~dbj}{http://star.pst.qub.ac.uk/$\sim$dbj}.}. The use of well-documented flare energy values naturally requires the implementation of a conversion mechanism to directly relate (nano)flare energetics to physical lightcurve intensity fluctuations. However, in order to maximize the usefulness of the Monte Carlo code to datasets obtained from a vast assortment of ground- and space-based instruments, the code does not rely on the error prone stipulation of crucial atmospheric (e.g., wavelength dependent transmission profiles) and instrumental (e.g., telescope throughput, detector quantum efficiencies, etc.) profiles that directly affect the calibration process. Instead, as described in detail in Appendix~{\ref{Appendix_power-law}}, the user can estimate the (nano)flare energy corresponding to the noise threshold of the chosen instrument, which will then be used to calibrate the resulting synthesized time series. Hence, the lightcurves output by the code will already be in data number (DN) units, often equally labelled as `counts', which are synonymous with the data products found in flagship observatories such as SDO. Such DN values are not just applicable to space-based observatories, but also to leading ground-based imaging and spectroscopic science data products, such as those from the Dunn Solar Telescope \citep[e.g.,][]{Cav06, Jaeggli10, Jes10b, Jess12} and the Swedish Solar Telescope \citep[e.g.,][]{Sch08}.

\subsection{Input Parameters selected for the Current Study}
In order to ensure the synthetic time series produced by the Monte Carlo code are representative of, and comparable with the SDO/AIA active region core, the time-averaged 171{\,}{\AA} and 94{\,}{\AA} snapshots (upper-left panels of Figures~{\ref{SDO171_images}} \&~{\ref{SDO94_images}}) formed the base science input images. For each simulation run, the base 171{\,}{\AA} or 94{\,}{\AA} science image defines the background count rates for each synthesized pixel, ensuring the reconstructed intensity fluctuation histograms from the Monte Carlo simulations are consistent with those generated directly from the SDO/AIA time series (see Appendix~{\ref{Appendix_power-law}} for more in-depth information). Furthermore, instead of assuming that the noise fluctuations embedded within the SDO/AIA lightcurves are purely photon based, we generated dedicated noise estimates following the methods detailed by \citet{Kirk14a, Kirk14b}, which are visible in the upper-right panels of Figures~{\ref{SDO171_images}} \& {\ref{SDO94_images}} and discussed in detail below. 

\subsubsection{SDO/AIA Noise Modeling}
Many techniques exist for estimating and removing noise from images. First, the point spread functions (PSFs) of the SDO/AIA imaging channels were generated following the methods detailed by \citet{2013ApJ...765..144P}, which are commonly available within standard {\sc{sswidl}} preparation routines. Once generated, the PSFs were deconvolved from the corresponding images using the standard and well-documented Richardson-Lucy algorithm \citep{Richardson:72, 1974AJ.....79..745L}. Such deconvolution assists with the removal of fringes that are caused by bright active regions, but also has the effect of modifying the fundamental statistical distribution of the residual noise. To account for this, we assume a model for calibration errors, instrumental resolution effects and compression artifacts as a combination of Gaussian and Poisson distributions. This is a reasonable assumption for the imaging detectors utilized by SDO/AIA, including the use of Rice compression that is often applied when transmitting the data \citep{2009PASP..121..414P}. 

Denoising images affected by Poisson noise is commonly performed by first applying a variance stabilizing transformation (VST) to standardize the image noise, then denoising the image using an additive white Gaussian noise filter, before returning the image to its original range via an inverse transformation \citep[e.g.,][]{2016ISPL...23.1086A}. In this work we used the same procedures. However, in the case of images with Poisson-Gaussian noise, such as with SDO/AIA \citep{2016JSWSC...6A...1G}, the generalized Anscombe transformation was used for stabilizing the noise variance \citep{Starck:1998:IPD:289385, 6212354}. This transformation generalizes the classical VST (i.e., the Anscombe transformation), which was designed for a purely Poisson noise mixture \citep{10.2307/2332343}. 

For SDO/AIA images, the high resolution and large number of pixels per image provides an advantage over other solar imaging platforms. We exploited the large field of view of our full-disk SDO/AIA time series to estimate remaining noise through the application of a non-local estimation technique. Termed ``block-matching'', the noise present in a small region of an image is estimated from other locations that are found to be statistically similar to the region of interest \citep{Buades:2005:NAI:1068508.1069066}. Block-matching methods of denoising, unlike transform-based ones such as wavelet denoising, introduce very few artifacts in the resulting estimates. We employed the BM3D \citep{2006SPIE.6064..354D} block-matching algorithm because of its high prevalence and good characterization abilities in signal processing communities. Preliminary tests \citep[see, e.g.,][]{Kirk14a, Kirk14b} found BM3D to be stable over a large range of intensities, which is important for examining the active region in the present study. Hence, we employed BM3D, an iterative block-matching routine with hard noise cutoff thresholds of the image in the sparse domain, to estimate the remaining noise in the SDO/AIA time series.  

Importantly, it must be noted that the spatial structuring present in the derived noise images map very closely to their corresponding time-averaged 171{\,}{\AA} or 94{\,}{\AA} snapshots, verifying that the Poisson statistics associated with shot noise distributions remains consistent with photon noise being a dominant source of error in SDO/AIA intensity measurements \citep[see also][]{DeForest2017}.

\subsubsection{Additional Model Parameter Definitions}
For the synthesis of subsequent SDO/AIA 171{\,}{\AA} and 94{\,}{\AA} time series, we employed a cadence of 12{\,}s, and due to the SDO/AIA pixel size of $0.6{\arcsec} \times 0.6{\arcsec}$, we set the surface area per synthetic pixel equal to $1.89225 \times 10^{15}${\,}cm$^{2}$ to accurately reflect the SDO/AIA platescale. In order to cover the vast assortment of suggested power-law indices, a grid of 51 indices spanning $1.5 \leq \alpha \leq 2.5$ (in intervals of 0.02) were submitted as inputs. Similarly, to cover $e$-folding times spanning both chromospheric \citep[e.g., $\sim$51{\,}s;][]{Jess14} and coronal \citep[e.g., $\sim$$360-1000${\,}s;][]{Ter11, 2018ApJ...864....5M} values, we chose a grid of decay timescales, $\tau$, equal to ${\lbrack}10,$ $20,$ $30,$ $40,$ $\cdots,$ $1000{\rbrack}${\,}s. A wide range of $e$-folding times have been chosen to span the entire spectrum reported in the literature, which include values as short as $\sim$50{\,}s \citep{Simoes2015}, through to long-duration ($\sim$1000{\,}s) decay times linked to low-frequency nanoflare models \citep{2018ApJ...864....5M}. Importantly, \citet{2008ApJ...677.1385C} have shown that decay timescales are not always constant between successive flaring events. As such, the $e$-folding time, $\tau$, for each modeled impulsive event is allowed to vary randomly (yet following a normal distribution centered on the relevant mean) by $\pm10$\%, which allows for some fluctuation in the specific decay times as a result of varying quiescent plasma parameters. The physics of a cooling plasma will not necessarily follow an exponential decay, but a confined spread (i.e., $\pm10$\%) of decay times will help cover small-scale permutations in the mechanisms that govern the rates of evaporative, non-evaporative, conductive and radiative cooling processes \citep{Ant78}. Since we are primarily concerned with intensity fluctuations arising as a result of small-scale (nano)flare events, we restricted the range of energies simulated to $10^{22}\le{E}\le10^{25}$~erg, where the resulting frequency distribution (for a power-law index of $\alpha = 2.24$) can be viewed in the upper-left panel of Figure~{\ref{flare_freq_dist}}. Finally, the flare energy corresponding to an average 1{\,}$\sigma_{N}$ intensity fluctuation was set as $5\times10^{24}$~erg, meaning that flare energies in the range of $5\times10^{24}<{E}\le10^{25}$~erg will exhibit intensity fluctuations greater than 1{\,}$\sigma_{N}$, while all other flaring energies will be represented by intensity fluctuations that are contained within the standardized noise envelope. 

Following the analysis of \citet{Lopez2016}, background intensities spanning a range of 120{\,}--{\,}5700~DN, which is synonymous with the present SDO/AIA active region core observations (see, e.g., the upper-left panels of Figures~{\ref{SDO171_images}} \&~{\ref{SDO94_images}}), may be expected to have standard deviations of approximately 10{\,}--{\,}1140~DN, which is consistent with the maps displayed in the lower-left panels of Figures~{\ref{SDO171_images}} \&~{\ref{SDO94_images}}. \citet{Price2015} have shown that nanoflare energies $\sim$$2\times10^{24}$~erg may demonstrate, when passed through appropriate forward modeling software \citep[e.g.,][]{Bradshaw2011}, SDO/AIA 171{\,}{\AA} intensity fluctuations on the order of 300{\,}--{\,}600~DN, which is in the middle-to-lower end of the standard deviation range derived following the work of \citet{Lopez2016}. Therefore, the (nano)flare energy corresponding to the full magnitude of a time series standard deviation (i.e., 1{\,}$\sigma_{N}$) is likely to be higher than the $\sim$$2\times10^{24}$~erg modeled by \citet{Price2015}. Hence, we have adopted a (nano)flare energy of $5\times10^{24}$~erg that corresponds to an average 1{\,}$\sigma_{N}$ intensity fluctuation in the SDO/AIA 171{\,}{\AA} synthetic time series, which is consistent with the minimum thermal energy ($\sim$$7\times10^{24}$~erg) predicted by \citet{2013ApJ...776..132A} to provide measurable nanoflare signatures. However, neither \citet{Price2015} nor \citet{Lopez2016} investigated the SDO/AIA 94{\,}{\AA} channel. Therefore, for consistency, we adopt the same 1{\,}$\sigma_{N} = 5\times10^{24}$~erg condition for the generation of synthetic SDO/AIA 94{\,}{\AA} time series. All input parameters employed for the current study are listed in Table~{\ref{tab:table1}}.

\subsection{Nanoflare energy and SDO/AIA Intensity Scaling}
\label{nanoflare_equations}
An important aspect to consider is how the SDO/AIA observables (i.e., the pixel count rates) scale with the magnitude of a flaring event. Due to the fact nanoflares are small-scale (in terms of their relative individual total energies) and presently unresolvable by current imaging instruments, there is no direct evidence available to say whether they extend beyond the $\sim$$190{\,}000$~km$^{2}$ spatial scales captured by an SDO/AIA pixel. Furthermore, since the power-law index governs the amount of flaring events per SDO/AIA pixel (i.e., $dN/dE$), any leakage of nanoflare emission between pixels would need to be negated by a reduction in flux from a neighboring region in order to satisfy a particular global power-law index, $\alpha$. Thus, as a first-order approximation, we assume a single nanoflare event occurs in a magnetic strand that is one SDO/AIA pixel wide. In these more simplistic terms, the total energy provided by a nanoflare contributes directly to intensity fluctuations captured by the SDO/AIA detectors. However, determining how different nanoflare energetics contribute to measurable intensity fluctuations requires a more in-depth examination of the inherent flaring plasma relationships. 

Although loop scaling laws \citep{1978ApJ...222..317R} were derived for loops at equilibrium between energy input and losses by radiation and conduction, \citet{2007A&A...471..271R} has shown that the equilibrium temperature is reached soon after the heating pulse commences, and that cooling starts immediately once the heating pulse finishes. According to the scaling laws, the maximum loop temperature, $T$, is defined by,
\begin{equation}
T = a(pL)^{1/3} \ ,
\label{eq:rtv1}
\end{equation}
where $a$ is a constant, $p$ is the pressure and $L$ is the loop half-length. The heating rate per unit volume, $H$, is given by,
\begin{equation}
H = b p^{7/6} L^{-5/6} \sim b \frac{p}{L} \ ,
\label{eq:rtv2}
\end{equation}
where $b$ is a constant. Combining Equations~\ref{eq:rtv1} \& \ref{eq:rtv2}, it is possible to further simplify to,
\begin{equation}
H = c \frac{T^{3}}{L^{2}} \ ,
\label{eq:rtvht}
\end{equation}
where $c$ is a constant. The accumulated SDO/AIA detector counts (DN) per pixel can be given by,
\begin{equation}
{\mathrm{DN}} = G n^{2} \Delta z \ ,
\label{eq:dn1}
\end{equation}
where $n$ is the plasma density, $\Delta z$ is the thickness of the emitting plasma along the line of sight, and $G$ is the SDO/AIA channel sensitivity per unit emission measure \citep[similar to, e.g.,][]{2009A&A...494.1127R}. Here, we assume that the SDO/AIA channels detect the cooling plasma predominantly at the temperatures where they demonstrate maximum sensitivity. Therefore, for the purpose of scaling, we assume for $G$ its maximum value. From the equation of state,
\begin{equation}
p = 2 n k_{B} T \ ,
\end{equation}
where $k_B$ is the Boltzmann constant, under the assumption that the maximum loop density is not far from that at equilibrium \citep{2007A&A...471..271R}, it is possible to combine this with Equation~\ref{eq:rtv1} to obtain,
\begin{equation}
n^{2} = d \frac{T^{4}}{L^{2}} \ ,
\end{equation}
where $d$ is a constant \citep{2007A&A...471..271R}. Combining with Equation~\ref{eq:rtvht}, we obtain,
\begin{equation}
T^{4}=f H^{4/3} L^{4/3} \ ,
\end{equation}
and,
\begin{equation}
n^{2}= g \frac{H^{4/3}}{L^{2/3}} \ ,
\end{equation}
where $f$ and $g$ are constants. Therefore, if we assume that the density decreases much more slowly than the temperature \citep{2007A&A...471..271R}, then the detector counts, DN, measured by an SDO/AIA detector when the cooling plasma crosses its temperature sensitivity can be written as,
\begin{equation}
{\mathrm{DN}} = h \frac{H^{4/3}}{L^{2/3}} \Delta z \ ,
\end{equation}
where, again, $H$ is the heating rate per unit volume, $L$ is the loop half-length, $\Delta z$ is the thickness of the emitting plasma along the line of sight, and $h$ is a constant. Classically, the injected nanoflare energy, $E$, can be written as,
\begin{equation}
E = H \Delta t \Delta V \ ,
\end{equation}
where $\Delta t$ and $\Delta V$ are the duration and encompassing volume, respectively, of the nanoflare heating pulse. \citet{2013ApJ...770..126B} have shown that the total radiated energy of a flaring event is proportional to the radiation produced during the rise phase alone. As such, the impulsive rise of the flare is the most crucial component when determining the relationship between flare energy and the resulting intensity fluctuation. In addition, \citet{2005ApJ...633..489R} have demonstrated through one-dimensional hydrodynamic modeling that nanoflare heating can provide temperature enhancements on the order of $1-1.5$~MK, which is well within the response functions of the SDO/AIA channels \citep{Boerner2012}. Furthermore, storms of nanoflares with relatively small ranges of energies and durations have been shown to accurately reproduce single-pixel loop lightcurves in the SDO/AIA imaging bands \citep{Taj16a}. We therefore do not expect that the volume, $\Delta{V}$ (in which the energy is produced), to change much from one event to another. Thus, for our estimates we can assume that the length of the strands, $L$, and the thickness of the emitting plasma along the line of sight, $\Delta{z}$, also remain relatively constant within the same active region.

\begin{figure*}
\begin{center}
\includegraphics[width=0.83\textwidth, clip=true]{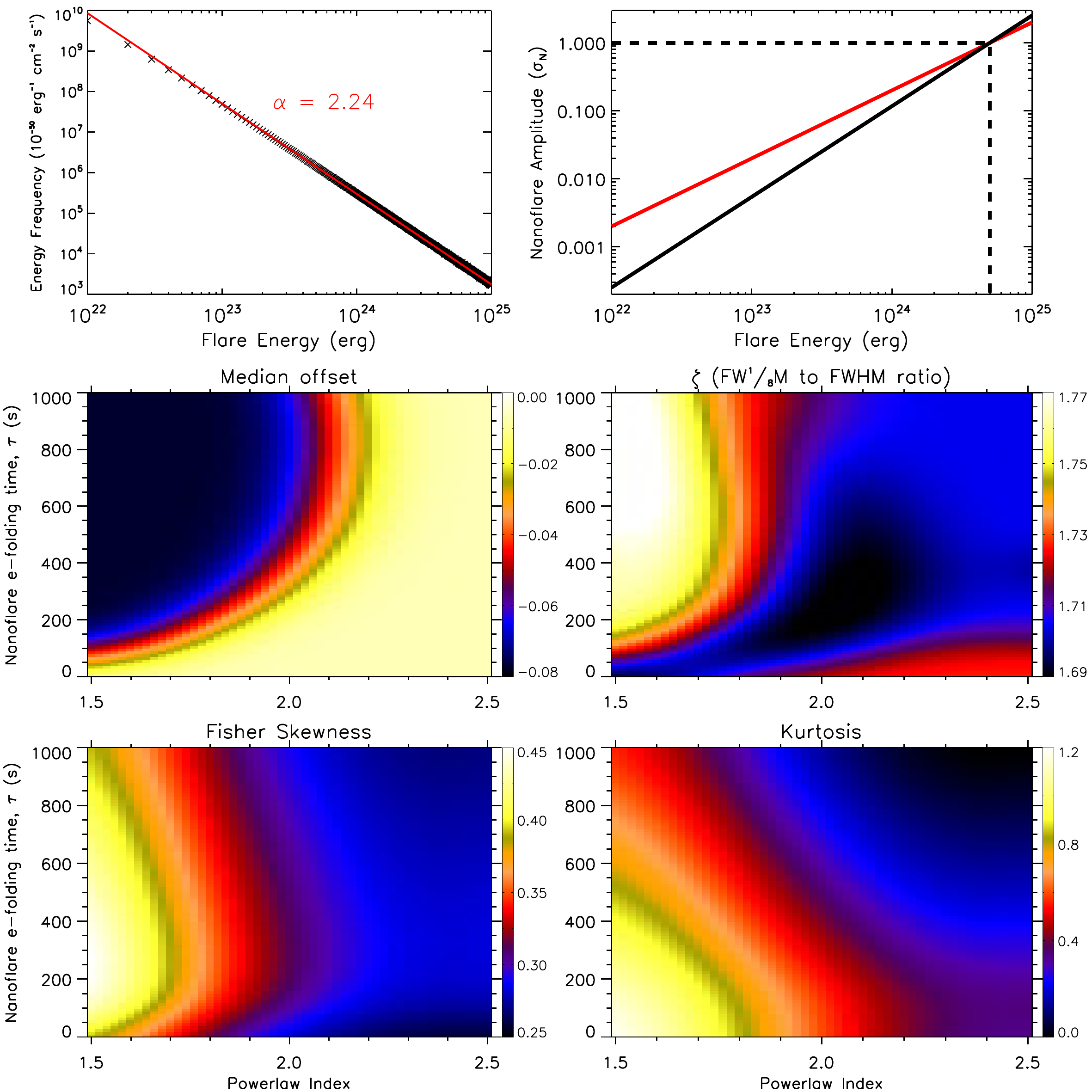}
\end{center}
\caption{A flare energy frequency distribution following a power-law index $\alpha = 2.24$ (upper-left), where the occurrence of flaring events (in units of $10^{-50}${\,}erg$^{-1}${\,}cm$^{-2}${\,}s$^{-1}$) is displayed as a function of the flare energy (in units of erg) using a log--log scale. The upper-right panel depicts the derived relationship between the nanoflare energy and the resulting simulated intensity amplitude (solid black line), which is given by ${\mathrm{DN}} \propto E^{4/3}$ (see Equation~{\ref{eq:final}}) and subsequently normalized to the background noise level, $\sigma_{N}$, for a nanoflare energy equal to $5\times10^{24}$~erg. A solid red line is drawn for comparison that indicates a linear (1:1) relationship between nanoflare energies and the corresponding intensity fluctuations, which overestimates and underestimates the true intensity perturbations at lower and higher nanoflare energies, respectively. The dashed black line highlights the background noise normalization process, whereby a $1{\,}\sigma_{N}$ intensity fluctuation equates to a $5\times10^{24}$~erg nanoflare. The lower four panels display the median offsets, FW$\frac{1}{8}$M-to-FWHM (i.e., $\zeta$) ratios, Fisher skewness and kurtosis characteristics as a function of the nanoflare $e$-folding time, $\tau$, and the power-law index, $\alpha$, used to generate the synthetic time series (see, e.g., Table~{\ref{tab:table1}}).
}
\label{flare_freq_dist}
\end{figure*}

To a first-order approximation, this provides a relationship between the peak detector counts, DN, and the energy, $E$, of the injected nanoflare events equal to,
\begin{equation}
{\mathrm{DN}} \propto E^{4/3} \ .
\label{eq:final}
\end{equation}
Hence, doubling the nanoflare energy will result in $\approx$$2.5$ times the peak intensity fluctuation captured by the SDO/AIA imaging detectors. For example, a baseline $5\times10^{24}$~erg nanoflare will correspond to an intensity fluctuation of $1{\,}\sigma_{N}$ (see the definition made in Table~{\ref{tab:table1}}), while a nanoflare of energy $1\times10^{25}$~erg will provide an intensity perturbation equal to $\approx$$2.5{\,}\sigma_{N}$. The relationship between the injected nanoflare energies and the resulting peak intensity fluctuations is documented in the upper-right panel of Figure~{\ref{flare_freq_dist}}, which is used in the Monte Carlo software to convert the randomized power-law nanoflare energies into modeled intensity perturbations. 

It must be noted that the presented energy/intensity scaling law is for a plasma that is pulse-heated, i.e., the plasma gets heated rapidly (instantaneously in the case of our Monte Carlo simulations) to its maximum temperature and then allowed to cool. Thus, each SDO/AIA channel captures the plasma during the cooling phase, which occurs as soon as it crosses the (relatively narrow) temperature range that it is sensitive to. This is especially true for the 171{\,}{\AA} channel, but also for the hot peak of the 94{\,}{\AA} channel if the heat pulse is sufficiently strong. In this scenario, the plasma density is assumed not to change much during the fast plasma cooling phase, and is the important parameter when determining the corresponding SDO/AIA emission when the plasma becomes visible in the selected channel (i.e., intensity is proportional to the square of the density in an optically thin plasma). This density is, in turn, determined exclusively by the magnitude of the heat pulse \citep[see, e.g., the discussion in][]{Cargill20140260}.

\subsection{Simulation Outputs}
Only two graphical axes are required to visualize the statistical parameter space provided by the Monte Carlo simulation outputs, corresponding to the power-law index and the $e$-folding timescale. This is due to the power-law index containing information related to both the energy-based intensity fluctuations and the frequency-based occurrence distributions. As a result, it is straightforward to display the statistical relationships arising from the interplay between the power-law indices and the decay timescales. The lower panels of Figure~{\ref{flare_freq_dist}} depict the statistical outputs of the Monte Carlo simulations as a function of the nanoflare $e$-folding timescale, $\tau$, and the power-law index. The variations in the parameters are quite remarkable, with significant trends visible in each of the median offset, FW$\frac{1}{8}$M-to-FWHM ratio (here defined as `$\zeta$' for simplicity), Fisher skewness and kurtosis plots. 

As documented by both \citet{Ter11} and \citet{Jess14}, the median offset increases to larger negative values with increasing fractions of higher-energy nanoflares (i.e., smaller power-law indices). A more energetic nanoflare provides a larger intensity amplitude, which results directly in a rise in the mean intensity of the lightcurve. However, the relatively rapid exponential decreases in the lightcurve intensities increases the separation between the statistical mean and median, hence increasing the magnitude of the median offset. It must also be noted that under no circumstances does the median offset become positive (middle-left panel of Figure~{\ref{flare_freq_dist}}), an indication that all possible permutations demonstrate some aspect of impulsiveness followed by rapid (i.e., exponential) decay. Such lightcurve shape is consistent with previous flare observations \citep[e.g.,][]{Qiu2012}, including those close to the nanoflare energy regime \citep[e.g.,][]{Ter11}, and therefore reiterates the usefulness of the statistical median offset as a proxy for asymmetric behavior trapped within the background noise. From Figure~{\ref{flare_freq_dist}} it is clear that time series less heavily dominated by small-scale energetics (i.e., lower power-law indices resulting in less low-energy nanoflares) display more significant median offsets. This is due to the larger flare energy inputs being more isolated within the synthetic lightcurves, resulting in their associated intensity fluctuations becoming more pronounced in the subsequent statistical distributions, hence inducing a larger negative median offset. Due to the rapid changes in the median offset around a power-law index $\sim$2, this statistical parameter is a significant marker for determining the prevalence of small-energy nanoflares in observational time series.

The $\zeta$ ratio map displayed in the middle-right panel of Figure~{\ref{flare_freq_dist}} reveals a bifurcated trend as both the nanoflare $e$-folding time and power-law index are increased. A standardized Gaussian distribution has a natural ratio of $\zeta=1.73$, so from the middle-right panel of Figure~{\ref{flare_freq_dist}} it is clear that both narrower (i.e., $\zeta<1.73$) and wider (i.e., $\zeta>1.73$) widths are present depending on the values of the input parameters. In general, increases in the nanoflare decay timescales result in broader tails of the corresponding intensity fluctuation histograms (i.e., $\zeta>1.73$). Such increased widths are a direct consequence of the injected nanoflare amplitudes being coupled with larger decay timescales, which allows the inclusion of significantly high positive $\sigma_{N}$ intensity fluctuations in the resulting statistical distributions, thus providing high values of the $\zeta$ ratio. The bifurcated signal displayed in the middle-right panel of Figure~{\ref{flare_freq_dist}} shows reduced distribution tail widths in the presence of increasing nanoflare decay timescales and large power-law indices. Now, rapidly injected impulsive events (i.e., corresponding to large power-law indices), provides the superposition of new nanoflare signals existing on top of previously decaying signatures. This reduces the contributions from large $\sigma_{N}$ intensity fluctuations (both negative and positive), which causes the tails of the statistical distributions to be pulled inwards, thus making them appear narrower than a standardized Gaussian (i.e., $\zeta<1.73$). Of course, very rapid $e$-folding timescales negate this effect since the rapid disappearance of previously injected nanoflares alleviates the continued superposition between new and existing nanoflare signals, hence the $\zeta$ ratio at locations of low $e$-folding times and larger power-law indices returns to standardized values.

\begin{figure*}
\begin{center}
\includegraphics[width=0.8\textwidth, clip=true]{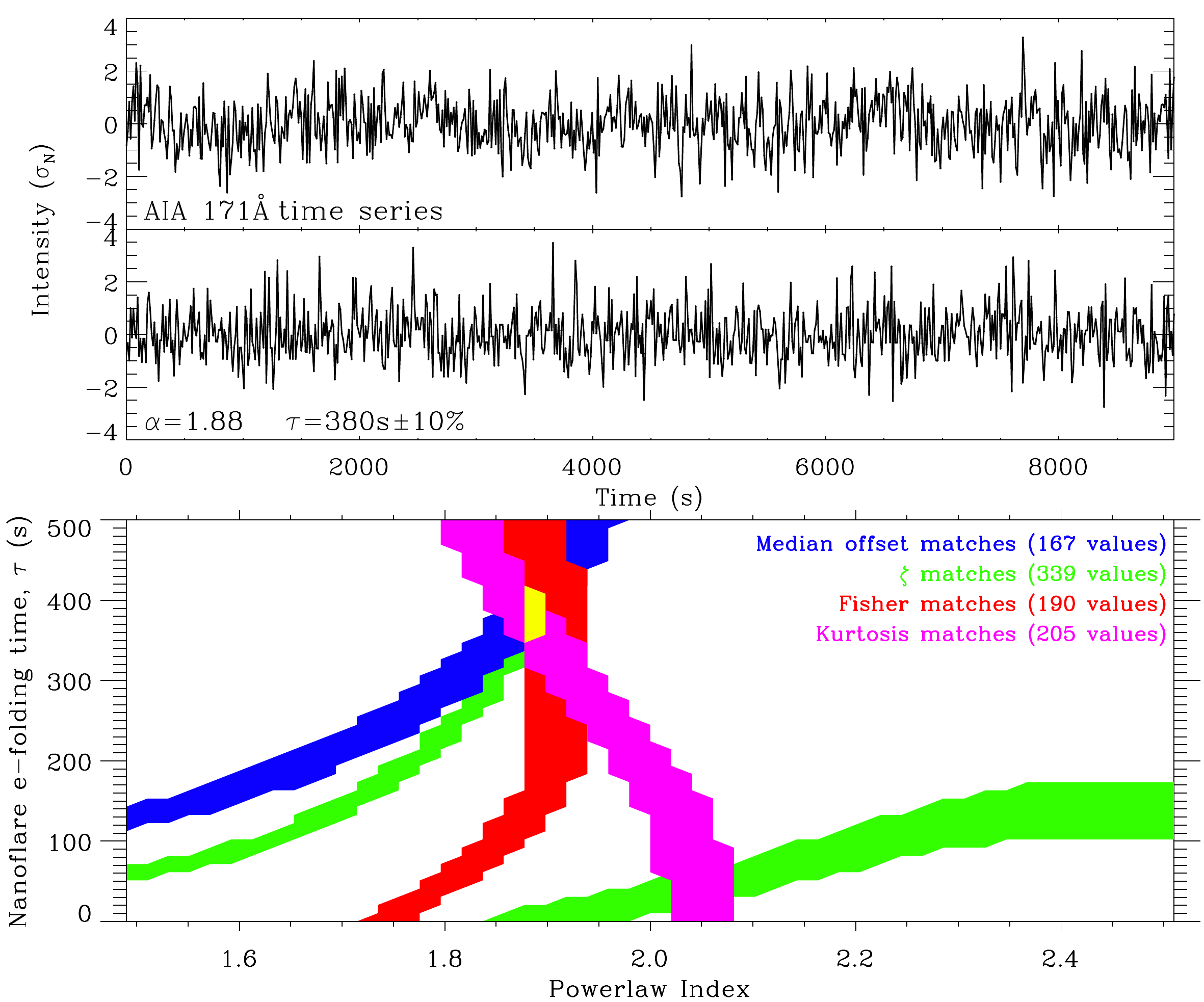}
\end{center}
\caption{A time series extracted from the SDO/AIA~171{\,}{\AA} dataset (top), contrasted with a synthetic power-law-based Monte Carlo lightcurve for a power-law index of $\alpha = 1.88$, coupled with a decay timescale of $\tau=380${\,}s~$\pm{\,}10$\% (middle). Both time series are normalized to the standard deviation of their respective noise ($\sigma_{N}$). The lower panel contours the matching statistical parameters obtained from the Monte Carlo simulations when compared to the SDO/AIA~171{\,}{\AA} observational distributions, with the decay timescale range cropped between $0 - 500${\,}s for clarity. The blue, green, red and pink contours representing the median offset, $\zeta$ ratio, Fisher skewness and kurtosis matches, respectively. Yellow contours indicate locations where the Monte Carlo outputs provide overlap of all possible statistical parameters.
}
\label{power-law_relationships_SDO171}
\end{figure*}

The Fisher skewness of the intensity fluctuation distributions are increased as a result of progressively smaller power-law indices (lower-left panel of Figure~{\ref{flare_freq_dist}}). Since a reduction in the power-law index promotes more isolated nanoflare signals, this will result in larger measurable $\sigma_{N}$ intensity fluctuations, hence providing elevated (and therefore skewed) tails in the statistical distributions. Measurements of the kurtosis remain positive across all $e$-folding timescales and power-law indices, as demonstrated in the lower-right panel of Figure~{\ref{flare_freq_dist}}. Here, the leptokurtic (i.e., narrow) nature of the intensity fluctuation distributions, corresponding to small power-law indices, can easily be related to the prevalence of small-scale fluctuations resulting from the ubiquity of the lower-energy (and therefore lower $\sigma_{N}$) intensity perturbations. For diminishing power-law indices, these small-scale $\sigma_{N}$ fluctuations begin to dominate the resulting intensity fluctuation distributions, hence becoming statistically leptokurtic and promoting positive values of kurtosis. Importantly, the Fisher skewness and kurtosis values displayed in the lower panels of Figure~{\ref{flare_freq_dist}} demonstrate their importance as a diagnostic tool, especially since they document significant statistical changes as the power-law index approaches a value of 2.

The ultimate goal is to be able to identify key nanoflare characteristics in real observational data by comparing their statistical distributions with those of synthesized (and well-defined) nanoflare activity. In essence, the median offset, $\zeta$ ratio, Fisher skewness and kurtosis values measured from observations could be used to pinpoint the specific power-law index, $\alpha$, and nanoflare $e$-folding time, $\tau$, parameters that correspond to identical simulated signatures. Previously, \citet{Ter11} and \citet{Jess14} utilized a negative median offset as the sole determining parameter in the estimation of observational nanoflare characteristics. However, from the examination of Figure~\ref{flare_freq_dist}, it becomes clear that a level of ambiguity arises as a result of the same median offset value being produced from a range of different nanoflare attributes. Thus, employing not just one, but four (median offset, $\zeta$ ratio, Fisher skewness and kurtosis values) independently measured statistical signatures will help reduce this ambiguity since all four parameters must intersect their corresponding parameter space at the same $\alpha$ and $\tau$ values to remain self-consistent with one another. This benefit is further maximized by the fact that each statistical signature displays parameter maps (e.g., Figure~{\ref{flare_freq_dist}}) that do not follow identical shapes or trends as their other statistical counterparts. 

\subsection{Comparing the Monte Carlo Statistical Parameter Space with SDO/AIA Data}
\label{testing_data02}
The median offset, $\zeta$ ratio, Fisher skewness and kurtosis coefficients established in Section~{\ref{observations}} for the SDO/AIA 171{\,}{\AA} observations, including the fitting errors stipulated in the lower panel of Figure~{\ref{SDO171_images}}, are used as contour thresholds on each of the parameter space windows displayed in the bottom four panels of Figure~{\ref{flare_freq_dist}}. The extracted contours relating to the median offset (blue), $\zeta$ ratio (green), Fisher skewness (red) and kurtosis (pink) are plotted in the lower panel of Figure~{\ref{power-law_relationships_SDO171}}, with yellow contours depicting the regions where all four statistical parameters overlap. It can be seen from the lower panel of Figure~{\ref{power-law_relationships_SDO171}} that the number of matches for the median offset (blue contours; 167 values) is fewest, while the number of matches related to the $\zeta$ ratio (green contours; 339 values) is largest. This is a consequence of the errors associated with each statistical parameter, hinting that future studies, which employ even more significant number statistics, will be able to reduce the number of matching {\it{and}} overlapping parameter values extracted from the lower panels of Figure~{\ref{flare_freq_dist}}. Here, there are 9 independent overlapping matches, which suggests for these specific synthetic input parameters, the power-law-based Monte Carlo simulations accurately depict the statistical processes occurring within the SDO/AIA~171{\,}{\AA} time series. All viable input parameters are within the ranges of $1.88 \leq \alpha \leq 1.90$ and $354{\,}{\mathrm{s}} \leq \tau \leq 410{\,}{\mathrm{s}}$ (see Table~{\ref{tab:table2}} and the yellow contours in the lower panel of Figure~{\ref{power-law_relationships_SDO171}}), with average values of $\bar{\alpha}_{171} = 1.89\pm0.01$ and $\bar{\tau}_{171} = 385\pm26${\,}s. The typical nanoflare decay timescale of $\bar{\tau}_{171} = 385\pm26${\,}s is very similar to that put forward by \citet{Ter11}, even though their approach did not hinge upon the use of power-law-based synthetic models. 

A sample de-trended and normalized SDO/AIA~171{\,}{\AA} lightcurve is displayed in the upper panel of Figure~{\ref{power-law_relationships_SDO171}}. A synthetic power-law-based lightcurve, generated using a nanoflare $e$-folding timescale $\tau=380{\,}{\mathrm{s}}{\,}\pm10$\% and a power-law index $\alpha = 1.88$, is displayed in the middle panel of Figure~{\ref{power-law_relationships_SDO171}}. Both time series are remarkably similar to one another, highlighting that ({\sc{i}}) an average 1{\,}$\sigma_{N}$ intensity fluctuation likely corresponds to a nanoflare energy $\sim$$5\times10^{24}$~erg, ({\sc{ii}}) the noise distribution is accurately modeled by Poisson-based terms, and ({\sc{iii}}) the de-trending and normalization approaches applied to both real and synthetic lightcurves (see Equation~{\ref{eqn:inten_normalization}}) are able to produce visually similar time series that can be directly compared in statistically significant ways.

\begin{table}
\centering
\caption{Statistical parameters extracted from the SDO/AIA~171{\,}{\AA} time series that intersect with the contour curves provided by the power-law-based Monte Carlo simulations depicted in the lower panel of Figure~{\ref{power-law_relationships_SDO171}}. Here, the power-law indices and $e$-folding timescales are most representative of the statistical parameter space deduced from the SDO/AIA~171{\,}{\AA} observations.}
\label{tab:table2}
\begin{tabular}{@{}lrr@{}}
\hline
{\bf{Output}}		& {\bf{Estimated power-law}}  		& {\bf{Estimated}}		\\
{\bf{matches}}	& {\bf{index ($\alpha$)}}				& {\bf{$\tau$ (s)}}		\\
\hline
1		& $1.88$						& 340	 \\
2		& $1.88$						& 350	 \\
3		& $1.88$						& 360	 \\
4		& $1.88$						& 370	 \\
5		& $1.88$						& 380	 \\
6		& $1.88$						& 390	 \\
7 		& $1.88$						& 400 \\
8		& $1.88$						& 410 \\
9		& $1.88$						& 420 \\
10	& $1.88$						& 430 \\
11		& $1.90$						& 360 \\
12	& $1.90$						& 370 \\
13	& $1.90$						& 380 \\
14	& $1.90$						& 390 \\
15	& $1.90$						& 400 \\
16	& $1.90$						& 410 \\
\hline
{\bf{Average value}}				& {\bf{1.89}}	& {\bf{385}} \\
{\bf{Standard deviation}}		& {\bf{0.01}}	& {\bf{26}} \\
\hline
\end{tabular}
\end{table}

\begin{figure*}
\begin{center}
\includegraphics[width=0.75\textwidth, clip=true]{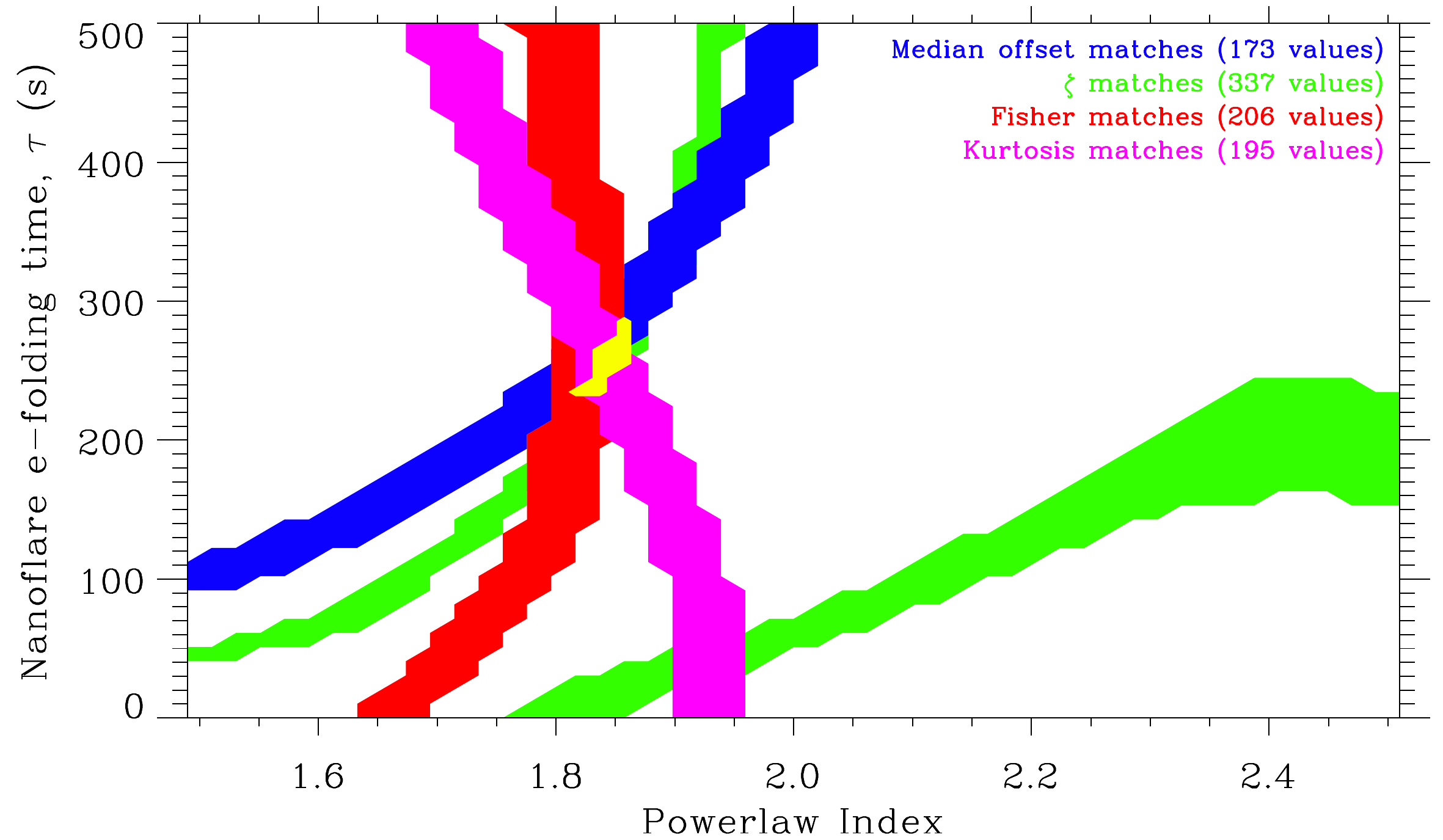}
\end{center}
\caption{A graphical representation identical to the lower panel of Figure~{\ref{power-law_relationships_SDO171}}, whereby the matching median offset, $\zeta$ ratio, Fisher skewness and kurtosis statistical parameters obtained from a comparison between the Monte Carlo simulations and the SDO/AIA~94{\,}{\AA} observational distributions are shown in blue, green, red and pink contours, respectively. As per the lower panel of Figure~{\ref{power-law_relationships_SDO171}}, yellow contours indicate locations where the Monte Carlo outputs provide overlap of all possible statistical parameters.
}
\label{power-law_relationships_SDO94}
\end{figure*}

Similarly, the median offset, $\zeta$ ratio, Fisher skewness and kurtosis coefficients established in Section~{\ref{observations}} for the SDO/AIA 94{\,}{\AA} observations were also investigated by contouring their respective values on each of the parameter space windows displayed in the bottom four panels of Figure~{\ref{flare_freq_dist}}. In an identical way to the bottom panel of Figure~{\ref{power-law_relationships_SDO171}}, the extracted contours relating to the median offset (blue), $\zeta$ ratio (green), Fisher skewness (red) and kurtosis (pink) are plotted in Figure~{\ref{power-law_relationships_SDO94}}, with yellow contours depicting the regions where all four statistical parameters overlap. From inspection of Figure~{\ref{power-law_relationships_SDO94}} and Table~{\ref{tab:table3}}, the input parameters for the Monte Carlo simulations providing viable overlapping values are within the ranges of $1.82 \leq \alpha \leq 1.86$ and $240{\,}{\mathrm{s}} \leq \tau \leq 290{\,}{\mathrm{s}}$ (see the yellow contours in the lower panel of Figure~{\ref{power-law_relationships_SDO94}}), with average values of $\bar{\alpha}_{94} = 1.85\pm0.02$ and $\bar{\tau}_{94} = 262\pm17${\,}s. 

\section{Discussion}
Following a comparison between the statistics linked to the intensity fluctuations of the SDO/AIA channels with those generated via our Monte Carlo nanoflare models, we find that the power-law index and $e$-folding timescale are both smaller for the 94{\,}{\AA} channel when compared to the 171{\,}{\AA} filter ($\bar{\alpha}_{\mathrm{94}} = 1.85\pm0.02$; $\bar{\tau}_{\mathrm{94}} = 262\pm17${\,}s versus $\bar{\alpha}_{\mathrm{171}} = 1.89\pm0.01$; $\bar{\tau}_{\mathrm{171}} = 385\pm26${\,}s). In terms of the smaller $e$-folding time associated with the SDO/AIA 94{\,}{\AA} channel, this may be a natural consequence of the hotter plasma temperatures sampled by this imaging filter. Reduced decay timescales for hotter channels have been demonstrated previously for observations and models of large-scale flaring events \citep[e.g.,][]{Petkaki2012, Qiu2013, Cadavid2014}. Indeed, \citet{Simoes2015} found short $e$-folding timescales on the order of $50$~s for very hot ($\sim$$13$~MK) post-flare plasma. This may be a consequence of more significant thermal conduction at greater temperatures above the coronal background \citep{Battaglia2009}, particularly in the presence of weaker flares \citep{2016A&A...588A.116W}, up to the point of saturation \citep{Simoes2015}. However, on the other hand, \citet{McTiernan1993}, \citet{Jiang2006}, \citet{Li2012} and \citet{Wang2015} have shown evidence for the suppression of thermal conduction in the vicinity of large-scale flares. As a result, if thermal conduction has the ability to be suppressed at small (nanoflare) energies, then radiative and/or collisional cooling may play an important role. Utilizing EUV observations obtained with the TRACE satellite, \citet{Aschwanden2000} provided evidence that the geometric and physical properties of EUV nanoflares represent miniature versions of larger-scale flare processes. This has important implications, especially since \citet{Aschwanden2000} suggest that nanoflares can be characterized by much smaller spatial scales and rapid heating phases, something that agrees with our modeling hypotheses put forward in Section~{\ref{nanoflare_equations}}. However, \citet{Aschwanden2000} also highlight that while the observed cooling times are compatible with theoretically calculated radiative cooling timescales, the theoretically calculated conductive cooling times are significantly shorter, thus requiring either high-frequency heating cycles \citep[e.g., similar to that documented by][]{War11} or reduced temperature gradients between the flaring loop tops and their corresponding footpoints. Nevertheless, our results clearly indicate that, at least for nanoflares, the decay timescales for plasma sampled in cooler SDO/AIA imaging bands are smaller than those found in the hotter channels. Theoretical work is still required to address the specific roles of conductive and radiative cooling in post-flare plasma \citep[e.g.,][]{1995ApJ...439.1034C, 2018ApJ...857...99D}, particularly in a regime dominated by small-scale nanoflare events.

\begin{table}
\centering
\caption{Same as Table~{\ref{tab:table2}}, only here for the most representative statistical parameters extracted from the SDO/AIA~94{\,}{\AA} time series.}
\label{tab:table3}
\begin{tabular}{@{}lrr@{}}
\hline
{\bf{Output}}		& {\bf{Estimated power-law}}  		& {\bf{Estimated}}		\\
{\bf{matches}}	& {\bf{index ($\alpha$)}}				& {\bf{$\tau$ (s)}}		\\
\hline
1		& $1.82$						& 240	 \\
2		& $1.84$						& 240	 \\
3		& $1.84$						& 250	 \\
4		& $1.84$						& 260	 \\
5		& $1.84$						& 270	 \\
6		& $1.86$						& 260	 \\
7 		& $1.86$						& 270 \\
8		& $1.86$						& 280 \\
9		& $1.86$						& 290 \\
\hline
{\bf{Average value}}				& {\bf{1.85}}	& {\bf{262}} \\
{\bf{Standard deviation}}		& {\bf{0.02}}	& {\bf{17}} \\
\hline
\end{tabular}
\end{table}

An important outcome of this work is the fact that a smaller power-law index of $\bar{\alpha}_{94} = 1.85\pm0.02$ is estimated for the SDO/AIA 94{\,}{\AA} time series, when compared to $\bar{\alpha}_{171} = 1.89\pm0.01$ for the co-spatial and co-temporal 171{\,}{\AA} observations. This effect may be related to the fraction of nanoflare events able to provide sufficient emission in the the hotter SDO/AIA 94{\,}{\AA} channel. If the weakest nanoflare events (e.g., those closest to the lower energy limit of $10^{22}$~erg) do not provide either sufficient thermalization of the plasma, or a large-enough filling factor of the corresponding SDO/AIA pixel, to manifest as detectable signal in the SDO/AIA 94{\,}{\AA} channel, then these low-energy events will be absent from the statistical parameters extracted from the observational time series. As a result, the estimated power-law index will be shifted to lower values, simply as a result of the hotter SDO/AIA 94{\,}{\AA} channel not adequately capturing the signal from the weakest of nanoflare events. This effect will become even more prevalent for regions that demonstrate an elevated excess of low-energy nanoflare events, e.g., $\alpha \gtrsim 2$.

The range of compatible power-law indices, for both the SDO/AIA 171{\,}{\AA} and 94{\,}{\AA} channels, are concentrated within the range of $1.82 \leq \alpha \leq 1.90$. This is consistent with \citet{Aschwanden2015}, who employed differential emission measure techniques on SDO/AIA observations to deduce multithermal energies that are consistent with RTV scaling laws \citep{1978ApJ...220..643R}. This suggests that, for this particular active region, nanoflares may not provide the dominant source of thermal energy in the corona; a natural consequence of $\alpha < 2$ \citep{Parker88, Hudson1991}. However, the active region studied for the present analysis is a decaying sunspot group, where the free magnetic energy responsible for reconnection phenomena may be substantially diminished. As a result, other heating processes (such as wave heating) may be responsible for the elevated temperatures found in the immediate vicinity of active region NOAA~11366.

Interestingly, the average power-law index found here is larger than that presented by \citet[][$\alpha \approx 1.35$]{Berg98}, who employed EUV observations from the Solar and Heliospheric Observatory \citep[SoHO;][]{1995SoPh..162....1D} to examine the occurrence rates and radiative losses of flares down to $\sim$$10^{24}$~erg. The work of \citet{Berg98} was in close agreement with the power-law index found by \citet[][$\alpha \sim 1.5$]{1995PASJ...47..251S}, who employed X-ray observations of transient coronal brightenings from Yohkoh \citep{Oga91} to examine reconnection events in the range of $10^{25} - 10^{29}$~erg. Importantly, the work documented here, alongside the findings of \citet{Berg98} and \citet{1995PASJ...47..251S}, provide evidence that small-scale ($10^{22} - 10^{29}$~erg) flaring events may not be the dominant source of thermalization in the solar corona. On the other hand, work by \citet[][$\alpha \approx 2.59$]{Krucker98}, \citet[][$\alpha \approx 2.52$]{Parnell00}, \citet[][$\alpha \approx 2.31$]{Benz02} and \citet[][$\alpha \approx 2.9$]{Wine02} provide observational evidence to the contrary, whereby the derived power-law indices suggest dominant heating by small-scale nanoflare activity. Of particular note is the work by \citet{Krucker98}, who examined emission measure fluctuations corresponding to small-scale reconnection events in the range of $8.0 \times 10^{24}$ to $1.6 \times 10^{26}$~erg that were captured by the EUV imager onboard SoHO. The authors found that at least $85\%$ of pixels within the field of view demonstrated significant fluctuations in their corresponding emission measures, which provided an estimated power-law index in the range of $2.3 < \alpha < 2.6$. With this power-law index in mind, \citet{Krucker98} estimate that at least 28{\,}000 reconnection events would need to take place across the whole Sun each second (down to an energy of $3 \times 10^{23}$~erg) in order to balance the known radiative losses. Such widespread coverage of pixels demonstrating statistical fluctuations \citep[i.e., the 85\% put forward by][]{Krucker98} is consistent with our global picture of NOAA~11366, whereby the fluctuation histograms depicted in Figures~{\ref{SDO171_images}} \& {\ref{SDO94_images}} correspond to $\sim$97\% and $\sim$86\% of the field of view for the SDO/AIA 171{\,}{\AA} and 94{\,}{\AA} channels, respectively.

Of course, these seemingly contradictory results (i.e., those documenting $\alpha < 2$ and those identifying $\alpha > 2$) are associated with a wide range of different active regions, sunspot types, parts of the solar cycle, wavelengths, resolutions, etc., which makes each measurement of the power-law index unique. As such, the observational evidence presented to date suggests that the local plasma conditions dictate whether nanoflare activity plays a dominant role in supplying thermal energy to maintain the multi-million degree coronal conditions. Establishing whether a global preference exists will require more statistical analyses across a number of solar cycles. 

Moving away from observational studies, there has also been a plethora of theoretical work undertaken to estimate the power-law index associated with flaring events. Recently, \citet{2018ApJ...854...14M} created a model whereby convective flows in granules force randomized motions to be generated at the footpoints of coronal loops, which produces twist in the magnetic fields, hence driving flaring events. The model of \citet{2018ApJ...854...14M} indicates an energy-dependent power-law index, whereby larger flares demonstrate shallower indices ($E \sim T^{1.0}$, where $T$ is the time interval between events) when compared to smaller energy events ($E \sim T^{1.5}$). This change in the power-law index as a function of flare energy has been witnessed in observations by a number of authors \citep[see, e.g., the review by][]{Aschwanden2016}. Employing a multi-threaded hydrodynamic simulation, \citet{2018ApJ...856..149R} forward-modeled the flare emission typically observed in Si~{\sc{iv}} and Fe~{\sc{xxi}} spectral lines. A power-law index of $\alpha = 1.5$ \citep[similar to that employed in the Monte Carlo models of][]{2009SoPh..255..211W} provided good agreement between the modeled Doppler shifts and event durations, when compared to those observed for the M-class flares captured by the Interface Region Imaging Spectrograph \citep[IRIS;][]{DeP14} spacecraft on 2015 March 12. Recently, \citet{2018arXiv180700763A} employed one-dimensional hydrodynamic models, alongside a three-dimensional reconstruction of an active region magnetic field, to examine how coronal temperatures and densities respond to nanoflare heating. The authors found that an energy power-law index of $\alpha = 2.4$ was required to bring modeled emission in-line with EUV spectroscopic observations. Hence, the range of power-law indices provided by modern theoretical work spans the values estimated in our present analysis. As such, more numerical (and statistical) modeling of nanoflare activity is required to see whether the diverse power-law indices currently predicted converge to more definitive values.

Additionally, our average power-law indices are slightly lower than those found for active stars in the range of $\alpha \sim 2.2-2.3$ \citep{Audard1999, Kas02}. For stellar cases, \citet{Arzner04} undertook an analytic approach to determine the amplitude distribution of flares that are commonly visible in the light curves from active stars. Comparing the computed histograms of counts and photon waiting times to real stellar flare distributions, \citet{Arzner04} found the best agreement with a relatively steep power-law index of $\alpha \sim 2.3$, implying small-scale reconnection events are important for the heating of stellar atmospheres. 

\section{Future Directions}
\label{future_directions}
The Monte Carlo methods presented here are suitable for modeling all atmospheric layers of the Sun. As a result, the work will naturally be applicable to ground-based observations of the upper photosphere and chromosphere using the, e.g., Rapid Oscillations in the Solar Atmosphere \citep[ROSA;][]{Jes10b}, Hydrogen-Alpha Rapid Dynamics camera \citep[HARDcam;][]{Jess12}, Interferometric BIdimensional Spectrometer \citep[IBIS;][]{Cav06} and CRisp Imaging SpectroPolarimeter \citep[CRISP;][]{Sch08} instruments. These simulations can also be directly compared to transition region observations using IRIS, in addition to broad-temperature optically thin coronal images acquired by, e.g., SDO/AIA and the upcoming Extreme UV Imager (EUI) onboard the Solar Orbiter \citep{Hal14}. Furthermore, \citet{Labonte2007} revealed that measurements of the variance in intensities from an X-ray source allows the mean energy per photon to be determined, hence allowing the methods presented here to be incorporated into similar photon spectroscopy studies.

As observational resolutions, particularly temporal cadences, become much higher, the `bottleneck' associated with achieving large number statistics no-longer arises as a result of a small field-of-view. This has important consequences, since it means that imaging datasets can be subdivided into regions of interest for further analysis, rather than being examined as an entire collective that may encompass a vast assortment of different solar features. Thus, a natural step would be to employ magnetograms of the photosphere to isolate various regions of distinct magnetism, including uni-polar, bi-polar and mixed polarity regions. Furthermore, utilizing magnetic field extrapolations \citep[e.g.,][]{Wiegelmann2004, Wiegelmann2007, Guo12, Aschwanden2013, Aschwanden2016b} and magnetohydrodynamic-based coronal magnetic field estimates \citep[e.g.,][]{Jess16} would allow the user to estimate the degree of non-potentiality in their preselected subfields, something that is often used as an indicator of the flaring capabilities of the plasma. Following these steps would allow the nanoflare characteristics associated with distinct solar structures and degrees of magnetic complexity to be investigated with high levels of precision.

A natural progression would be to employ multi-wavelength observations, spanning the radio domain \citep[e.g., ALMA;][]{Wedemeyer2015} through to X-rays \citep[e.g., Hinode/XRT;][]{Gol07}, to provide a multi-height investigation of nanoflare activity from the photosphere through to the outer corona. Even within the EUV environment sampled by SDO/AIA, the self-similar and complementary platescales and cadences may be utilized to examine the multi-wavelength, and hence multi-thermal effects associated with flaring events \citep[e.g., furthering the work of][]{Viall11, Viall12, Viall13, Viall15}. Here, the intensity fluctuations associated with nanoflare processes contained within the same field-of-view, yet spanning a multitude of different temperatures, can be probed, with the corresponding statistical parameter space examined to better understand the prevalence of nanoflares in the solar corona. Combining such observations with magnetic field extrapolations demonstrating low levels of field misalignment would not only highlight the atmospheric layers where nanoflare activity is most prevalent, but it would also provide crucial insight into the magnetic topologies required for heightened nanoflare signatures. For example, do nanoflares in the solar corona require increased magnetic field braiding to support small-scale reconnections in an inherently low plasma-$\beta$ (i.e., dominated by magnetic tension) environment? Similarly, do nanoflares in the chromosphere exist more readily in less magnetic locations where increased plasma pressure may promote the volume filling of the magnetic fields? These types of questions, among others, can be addressed through the comparison of our Monte Carlo simulations to multi-wavelength observations of high-cadence sub-fields.

The Monte Carlo nanoflare simulations themselves are readily suitable for future upgrades once new physical insight is uncovered following future comparisons with high-resolution observations. For example, since the noise levels in our simulated lightcurves are, by default, entirely comprised of Poisson-based shot noise, a natural question arises as to whether other noise sources may contribute to subtleties often displayed in the observational histograms. The intensity fluctuation histograms generated by \citet{Ter11} and \citet{Jess14} for pure shot noise closely follow a standardized Gaussian distribution, as one may expect in the limit of large number statistics. This simply means that shot noise may not be responsible for some of the offsets and asymmetries pertaining to our observational histograms. Instead, other less symmetric noise sources, especially in the realm of large number statistics, may be partly responsible for the visible variations. In particular, other types of noise such as Brownian or violet noise, which have a frequency-dependent amplitude, may play a role when attempting to model the quasi-periodic (i.e., {\it{almost}} frequency dependent) injection of nanoflares. Indeed, \citet{Ireland2015} and \citet{Milligan2017} recently performed Fourier analysis on SDO time series and revealed that the resulting power spectra may be comprised of both flaring power-law signatures and a wave-based leakage term linked to the underlying $p$-mode spectrum. Therefore, a natural piece of followup work would be to precisely characterize the noise contributions embedded within the observational data, and more precisely incorporate these into the Monte Carlo simulations. 

The layer of the solar atmosphere attempting to be quantified also has important implications on the underlying pixel intensities, and therefore, their intensity fluctuation distributions. For example, \citet{Law11} demonstrated how positive values of kurtosis may also suggest the presence of localized turbulence. As a result, turbulent motions occurring within the observed plasma layers may adversely affect the statistical analysis of the intensity fluctuations. In order to more accurately constrain the nanoflare parameters, it may become imperative to include turbulent modeling in the simulated lightcurves. Since modern day observatories and instrumentation are constantly improving the spatial, temporal and spectral resolutions available, small-scale turbulence, perhaps even down to the Kolmogorov length scales, may be necessary to model. 

Further, the methods and approaches presented here may also be useful in the studies of propagating disturbances (PDs) within coronal loops, fans and plumes \citep[e.g.,][to name but a few]{DeF98, Ofman99, DeM02, DeM03, DeM04, Klim04, Wang09, Jess12, Krishna15}. \citet{Wang13} employed simulations to indicate that PDs, which are ubiquitously observed in coronal SDO/AIA observations, may be produced by small-scale impulsive heating events (e.g., nanoflares) at the loop footpoints. However, the models employed by \citet{Wang13} utilize a more straightforward cosine-based impulsive functional form, rather than the traditional impulsive rise and gradual decay found in many small-scale flare studies. Furthermore, since \citet{Keys11} have also documented how the velocity evolution of small-amplitude flares closely matches the intensity fluctuations, the Monte Carlo based nanoflare time series presented here could equally be employed to model intensity or velocity impulses at the base of coronal loops. These more traditional impulsive rises and exponential decays could then be fed into magneto-hydrodynamic modeling codes, such as that used by \citet{Wang13}, to investigate in more detail whether nanoflares can contribute to the creation of PDs in coronal loops, fans and plumes. Additionally, \citet{Hudson2004} revealed evidence to support the earlier work of \citet{Uchida1968}, whereby wave motion in coronal loops can be associated with weakly non-linear effects originating in metric type~{\sc{ii}} bursts. Therefore, the inherent `saw-tooth' shape of the nanoflare lightcurves may be able to help probe some of the weakest oscillating structures, especially since chromospheric and coronal fluctuations are becoming increasingly ubiquitous with continual improvements in instrument spatial and temporal resolutions \citep{Jess2015, 2018NatPh..14..480G, 2018ApJ...860...28H}.

Finally, the nanoflare characterization approaches described here can also be directly applied to high time resolution observations of stellar sources \citep[e.g., modernizing the work of][]{Audard1999, Kas02}. For example, the photometric precision of high-cadence Kepler observations \citep{Koch10}, or those taken through an assortment of optical filters by ULTRACAM \citep{Dhi07} at very high cadences, would be ideal to test the nanoflare signatures on variable dwarf stars. Indeed, developing on from the work of \citet{Hawley14}, \citet{Pit14}, \citet{Balona15} and \citet{Lurie15}, the initial Monte Carlo simulations could be extended to larger amplitude values to expand the `nanoflare' coverage into more macroscopic sources that demonstrate intensity fluctuations exceeding $3{\,}\sigma_{N}$. With larger scale impulsive events being much more evident in the resulting lightcurves, the user would have a much better grasp of the $e$-folding timescales associated with the flaring activity. As a result, the occurrence rates of (quasi-)periodic activity could be much more reliably constrained, thus better revealing the energy output of such flaring events on distant stellar sources. In addition, high temporal resolution ($\sim$$0.1$~s) observations of the most active flare stars have revealed some extremely short duration events lasting for $5$~seconds or less. Unlike the nanoflare lightcurves examined here, these events often exhibit highly symmetrical lightcurves with very similar rise and decay times \citep{Andrews1989, Andrews1990a, Andrews1990b, Tov1997, Schmitt2016}. Such characteristics are likely to modify the statistical distributions depicted in the lower panels of Figure~{\ref{flare_freq_dist}}, since these events can no longer be defined by impulsive rise times. As a result, we plan to investigate such phenomena in detail in a future publication.

\section{Overview and Concluding Remarks}
The creation of dedicated Monte Carlo simulations, with dense input parameter grids, has been shown to be a powerful tool that may greatly assist with the characterization of nanoflare activity manifesting in observational datasets. In the present work we have documented statistical relationships as a function of the Monte Carlo input parameters, specifically the nanoflare power-law index, $\alpha$, and the $e$-folding times, $\tau$. We have shown how specific statistical measurements, including the shifts in the median offset, fluctuations in the degree of skewness, and variations of the histogram width (e.g., the $\zeta$ ratio), can result in significant ambiguities since identical values can be created through a wide range of different nanoflare power-law indices and $e$-folding timescales. Simply, this means that an observational dataset cannot be accurately quantified by comparing a single extracted statistical measurement to that output by a Monte Carlo simulation. Instead, the only way to accurately constrain the true observational makeup is to explore more-detailed parameter space and compare a wealth of statistical parameters simultaneously, including the median offset, the level of distribution skewness, the histogram widths (e.g., the $\zeta$ ratio) and the values of the kurtosis, to identical statistical measurements output by the simulations. Such relationships are applicable to all layers of the solar atmosphere, including the chromosphere and corona. We have demonstrated how our methods are suitable for comparison to current (and future) CCD/CMOS detectors with high dynamic ranges, including those that will be implemented on the upcoming $4${\,}m Daniel K. Inouye Solar Telescope \citep[DKIST, formerly the Advanced Technology Solar Telescope, ATST;][]{Kei03, Rim10}.

Through comparisons of the statistical outputs derived from the Monte Carlo nanoflare simulations with SDO/AIA EUV imaging observations, we have provided evidence that hotter SDO/AIA observations (e.g., from the 94{\,}{\AA} channel) demonstrate both smaller power-law indices and $e$-folding timescales than their cooler SDO/AIA imaging counterparts (e.g., the 171{\,}{\AA} channel). This may be a consequence of increased conductive cooling and fewer registered nanoflare signals (through, e.g., weaker thermalization for the small-energy nanoflare events) in the hotter ($>1$~MK) plasma captured by the SDO/AIA 94{\,}{\AA} filter. From the Figures presented in this work, it is clear that the statistical methods employed here are useful for accurately quantifying nanoflare characteristics manifesting in observational data. However, the next important step is to apply these techniques to a multitude of observational datasets spanning a wide range of wavelengths, atmospheric heights and magnetic field complexities, and attempt to probe the small-scale nanoflares therein that lie beneath the noise envelope. 

\acknowledgements
D.B.J. would like to thank the UK Science and Technology Facilities Council (STFC) for an Ernest Rutherford Fellowship, in addition to a dedicated standard grant which allowed this project to be undertaken. 
D.B.J., C.J.D., M.M. and S.D.T.G. also wish to thank Invest NI and Randox Laboratories Ltd. for the award of a Research \& Development Grant (059RDEN-1) that allowed the computational techniques employed to be developed.
D.J.C. would like to thank California State University Northridge for start-up funding. 
S.K.P. and M.M. wish to thank the UK STFC for support. 
The SDO/AIA imaging employed in this work are courtesy of NASA/SDO and the AIA, EVE, and HMI science teams. 

\appendix

\section{Monte Carlo simulations}
\label{Appendix_power-law}
\subsection{Code Overview and User Inputs}
\label{power-law_inputs}
As documented by \citet{Jess14}, the Monte Carlo nanoflare code is written in {\sc{idl}}, but is now parallelized to run simultaneously on all of the locally available CPU cores, as well as implementing a power-law-based occurrence of nanoflares with different energetics. The main task of the software is to generate a series of impulsive events, which are governed by individual decay timescales and occurrence rates, before superimposing these intensity signatures onto a background level and adding realistic noise characteristics to the resulting lightcurve. Thus, a synthetic time series is produced that is subsequently statistically analyzed. By employing a dense grid of input parameters that have underlying similarities with the observables under investigation (e.g., background count rates, noise levels, etc.), the outputs can then be compared statistically to the observables, whereby closely matching statistics allows the quantification of the underlying impulsive nature of the observations, which may not have been readily visible through direct inspection of the raw lightcurves due to small-scale fluctuations becoming swamped by instrumental and/or photon noise. 

A significant difference between the current Monte Carlo nanoflare code, and that first used by \citet{Jess14}, is the substitution of a flare energy frequency distribution in place of (nano)flare amplitudes and occurrence rates distributions. Since traditional flare energy power-law distributions contain information on both the flare energies (i.e., related to the amplitude of the time series fluctuations) as well as the occurrence rates at which they occur, the stipulated power-law index naturally replaces two of the initial variables required for the Monte Carlo code used by \citet{Jess14}, hence making it much more computationally efficient. The user has the option to supply a dense grid of input parameters that have underlying similarities with the observables under investigation, thus allowing the quantification of the underlying impulsive nature of the observations following comparison with the simulation outputs. Once the code has been initialized, a grid of input parameters can now be specified, allowing the program to continually process and output data without further operator input. Specifically, the input parameters required consist of:
\begin{enumerate}
\item{The quiescent background intensity, in data numbers (DN), on top of which impulsive signatures are injected. This is typically the time-averaged value of the observational region of interest, such as 4500{\,}DN in an SDO/AIA 171{\,}{\AA} active region core, or 28{\,}000{\,}DN for a ground-based 16-bit detector examining chromospheric plage.}
\item{The standard deviation of the noise level ($\sigma_{N}$), in units of DN, superimposed on top of the impulsive time series, which is assumed to be dominated by photon noise unless otherwise stipulated.}
\item{The dimensions of the dataset you wish to simulate, consisting of the $x$ and $y$ sizes, the number of successive frames, as well as the cadence of the final synthetic time series.}
\item{The flare energy power-law indices, $\alpha$, which are specified as a numerical value that follow the relation $dN/dE \sim E^{-\alpha}$. The input indices can cover a very large range, and thus span previous work documenting shallow \citep[e.g., $\alpha=1.35$;][]{Berg98} and steep \citep[e.g., $\alpha=2.90$;][]{Wine02} power-law distributions.}
\item{The nanoflare $e$-folding time(s), $\tau$, defined as the typical time frame in which the nanoflare intensity will decrease by a factor $1/e$.}
\item{The minimum and maximum flare energies wishing to be simulated, provided in units of erg. As a user may only wish to study the effect of small-scale impulsive events, they have the ability to stipulate an upper cut-off for the energetics simulated, while to minimize CPU run times the user may also wish to place a lower energy cut-off to reduce the number of simulated events to a range that becomes statistically significant.}
\item{The flare energy that corresponds to a 1{\,}${\sigma_{N}}$ intensity fluctuation. A user may feel that a highly sensitive imager has the ability to visually highlight $10^{25}${\,}erg events through readily apparent intensity rises above the quiescent (photon) noise level, either through predictive forward modeling or previous inspection. As a result, this value can be used to re-normalize the synthesized flare energies to ${\sigma_{N}}$ fluctuations for the purposes of statistical representation and analysis.}
\item{The surface area of one synthesized pixel, provided in units of cm$^{2}$. Since the area of the solar surface occupied by the synthetic time series (i.e., governed by the $x$ and $y$ dimensions, as well as the surface area of each pixel) will directly affect the number of flaring events expected, this value is important to accurately scale the flare energy frequency distribution to realistic values.}
\item{{\it{optional:}} A base science image that will form the initial starting conditions of the simulated time series. Here, the simulated $x$ and $y$ sizes will be identical to those from the input base image, and for each pixel iteration the background intensity, $n$, (and subsequent photon noise level, $\sqrt{n}$, unless an optional noise image [see below] is also submitted) will be extracted from the value provided by the base image. This base image could be, for example, a time-averaged cropped field-of-view from an SDO/AIA channel (see, e.g., the upper-left panels of Figures~{\ref{SDO171_images}} \&~{\ref{SDO94_images}}), allowing the final synthesized time series to be readily compared to the true observational data. Submission of a base science image overrides inputs 1{\,}--{\,}3.}
\item{{\it{optional:}} A base noise image that spatially represents the noise complexion of the image sequence wishing to be simulated. This image should be comprised of the same pixel dimensions as the base science image, with each pixel representing the standard deviation of the noise as measured from the original time series (see, e.g., the upper-right panels of Figures~{\ref{SDO171_images}} \&~{\ref{SDO94_images}}). If only a base science image is submitted, then by default, a synthetic noise image will be generated that mimics Poisson-based photon statistics (i.e., $\sqrt{\mathrm{science~image}}$). Providing a noise reference image naturally overrides input 2.
}
\end{enumerate}

While these form an extensive list of input parameters, they allow the best possible tailoring of the Monte Carlo simulations to physical data. Of particular interest is the ability to stipulate optional base science and noise images, which means that fluctuations and noise levels can be automatically and individually tailored to specific pixels within the field-of-view. This is very useful for observations that contain a vast assortment of features and structures that give rise to diverse background intensities spanning many orders of magnitude. 

\subsection{Analysis and Output of the Numerical Code}
\label{Monte_analysis}
The first important aspect to consider when attempting to synthesize a time series is how many flare events one may expect to populate the final lightcurves. Of course, while a standardized power-law distribution of the form $dN/dE \sim E^{-\alpha}$ may provide the frequency relationship between one flare energy and the next, it unfortunately does not provide specific occurrence values. Therefore, to accurately calibrate the expected occurrences of the (nano)flare activity, the first step is to employ the user defined pixel sizes and time series duration to map the expected number of events. Previous work \citep[e.g.,][to name but a few]{Berg98, Krucker98, Aschwanden1999, Parnell00, Benz02, Wine02, Aschwanden2012, Aschwanden2014, Aschwanden2015} quantify the flare energy frequencies in units of erg$^{-1}${\,}cm$^{-2}${\,}s$^{-1}$ (see, e.g., the upper-left panel of Figure~{\ref{flare_freq_dist}}). Therefore, it is clear that the surface area simulated (i.e., cm$^{2}$) and the duration of the time series (i.e., seconds), play a pivotal role in the expected number of flaring events. 

\vspace{3mm}
As a result, the power-law-based Monte Carlo code takes the $x$ and $y$ dimensions of the simulated field-of-view and multiplies by the pixel area to calculate the total surface area being synthesized. Then, the user defined cadence and desired number of modeled frames are used to calculate the total time duration of the final dataset. From here, it is now possible to re-sample the flare frequency distribution into quantified occurrences of the various flare energies. However, an important point is {\it{at what energy}} does the normalization occur? From inspection of, e.g., Figure~10 in \citet{Aschwanden2000} and the review article by \citet{Aschwanden2016}, it is clear to see that different flare frequencies have been found for identical flare energy values. This issue is further exacerbated by the fact that different power-law indices naturally cause the divergence of flare frequencies across a range of energies. Importantly, since flare occurrence rates are more accurately quantified at higher energy values as a result of better defined observational measurements, there is a reduced spread of frequency values found at larger flare energies. As a result, and because this Monte Carlo code is designed with small-scale flaring events in mind, the normalization process is undertaken at a value of $10^{25}${\,}erg, which is believed to be on the upper-end of traditional nanoflare energetics. Here, we adopt the flaring frequency of $\sim$$2000\times10^{-50}${\,}erg$^{-1}${\,}cm$^{-2}${\,}s$^{-1}$ found by \citet{Parnell00} for an energy of $10^{25}${\,}erg. Thus, our power-law distributions are subsequently re-normalized to this value, allowing a quantified number of flaring events to be computed for the field-of-view and time series duration specified. 

\vspace{3mm}
Using the flare energy value that corresponds to a 1{\,}$\sigma_{N}$ intensity fluctuation (estimated by the user during the initialization sequence of the code), each flaring event can be normalized by the energy-dependent intensity fluctuation relationship provided by Equation~{\ref{eq:final}} (and subsequently displayed in the upper-right panel of Figure~{\ref{flare_freq_dist}}) to produce an intensity fluctuation amplitude in terms of $\sigma_{N}$. It would not be practical for the Monte Carlo software to select an energy value corresponding to a 1{\,}$\sigma_{N}$ intensity fluctuation on an automatic basis. Since this code is applicable to a wide range of atmospheric layers, solar features, input wavelengths (and associated photon fluxes), cadences, camera architectures and measurement units, it would not be reliable to pre-select an energy value for the user. Therefore, we allow the user to employ a-priori knowledge \citep[e.g., through forward modeling approaches;][]{Price2015, Viall15} or previous examination of similar lightcurves \citep[e.g.,][]{Kowalski2016}, to provide a $\sigma_{N}$ normalization factor that is useful for statistical purposes. 

\vspace{3mm}
The use of an optional base image (see, e.g., item 9 in Appendix~{\ref{power-law_inputs}}) modifies the injected intensity fluctuation process when compared with a more simplistic constant background value. As one would expect, for each pixel the background intensity, $n$, is simply extracted from the corresponding pixel in the base image, with (unless otherwise defined) the magnitude of the noise following photon noise statistics with a standard deviation equal to $\sqrt{n}$ \citep[for an in-depth overview of shot-noise processes see, e.g.,][]{Del08}. This results in many different background intensity values alongside an equally diverse assortment of Poisson noise profiles. Then, for example, if an impulsive event with an energy of $10^{24}${\,}erg was synthesized, it would contribute to the lightcurves in slightly different ways; perhaps making a clear brightening in a previously relatively dark region of the field-of-view, or remaining swamped by the higher natural photon noise fluctuations synonymous with a brighter solar feature. Thus,  in order to ensure that the injected intensity fluctuations are consistent with both the flare energy frequency distributions and what would be expected of the background pixel values, a baseline normalization is required. To do this, the spatially averaged intensity (i.e., $\bar{n}$) within the field-of-view provided by the base image is computed, with the user specified flare energy corresponding to a 1{\,}$\sigma_{N}$ intensity fluctuation calculated in relation to $\sqrt{\bar{n}}$ (again, assuming photon noise statistics). Alternatively, if the user also submits an independent noise image containing the pixel-by-pixel noise standard deviations, then the average of these standard deviations will be used when equating the chosen energy value to a 1{\,}$\sigma_{N}$ intensity fluctuation. It must be stressed that this form of normalization is undertaken as the Monte Carlo program makes no assumptions regarding what features within the simulated field-of-view will be more prone to (nano)flare activity. Therefore, before submitting the Monte Carlo processing call, it is important that the user identifies what types of structures they wish to investigate. It may be more advantageous, as well as less CPU intensive, if the user further crops the simulated field-of-view to more directly encompass the features they wish to simulate, since including a vast assortment of irrelevant structures with vastly differing background intensities may adversely affect the resulting simulated outputs.

\vspace{3mm}
While three-dimensional inputs ($x$ and $y$ sizes plus the number of consecutive frames) are fed into the numerical code, each time series is processed individually before finally being combined into an average statistical distribution. Following the creation of a power-law-based intensity distribution for the entire field-of-view, the code then subsequently computes additional distributions based upon the specified input parameters such that,
\begin{enumerate}
\item{Decay times for individual impulsive events are generated by computing a normal distribution, ranging from $90$\% to $110$\% of the selected $e$-folding time, $\tau$, which allows for some fluctuation in the specific decay times as a result of varying quiescent plasma parameters. The physics of a cooling plasma will not necessarily follow an exponential decay, but a confined spread ($\pm10$\%) of decay times will help cover small-scale permutations in the mechanisms that govern the rates of evaporative, non-evaporative, conductive and radiative cooling processes \citep{Ant78}.}
\item{Noise amplitudes for each time step are generated by creating a Poisson distribution centered on the value of the pre-defined noise, which without external input is deemed to tend to shot noise ({\rmfamily i.e.,} $\sqrt{n}$) in the limit of adequate light levels.}
\end{enumerate}

Next, using randomized values for each distribution, which are themselves governed by a constantly evolving seed function, we introduce a series of impulsive intensity rises on top of the quiescent background, followed by exponential decays, before superimposing noise fluctuations to replicate a typical time series embedded with nanoflare activity. Remaining consistent with the processing steps of observational time series, the time-resolved intensity fluctuations, $dI$, are computed in an identical fashion to Equation~{\ref{eqn:inten_normalization}},
\begin{equation*}
dI(t) = \frac{I(t) - I_{0}(t)}{\sigma_{N}} \ ,
\end{equation*}
where $I(t)$ and $I_{0}(t)$ are the count rate and value of a linear least-squares fit, respectively, at time $t$, and $\sigma_{N}$ is the magnitude of the noise superimposed on top of the time series, which in the limit of Poisson statistics ({\rmfamily i.e.,} for shot-noise dominated lightcurves) is approximately equal to the standard deviation of the normalized pixel lightcurve. 
The middle panel of Figure~{\ref{power-law_relationships_SDO171}} represents a synthetic nanoflare time series generated for relatively small-energy impulsive events. It is clear from Figure~{\ref{power-law_relationships_SDO171}} that small-scale nanoflare amplitudes are visually lost within the photon noise, which is consistent with the small amplitude and `low frequency' nanoflare scenario discussed by \citet{Car94} and \citet{Car97, Car04}. Poisson noise statistics will naturally introduce a degree of asymmetry to any photon-based signal distribution as a result of discrete data sampling. However, as documented by \citet{Ter11} and \citet{Jess14}, large-scale sample sizes result in the typically asymmetric Poisson distributions becoming more Gaussian-like, and hence more symmetric. Thus, when an operator selects increasingly large numbers of frames to simulate, any resulting asymmetries present in the statistical output cannot be directly attributed to Poisson noise statistics alone. 

\vspace{3mm}
A histogram of all $dI$ values is then computed, which by definition has a statistical mean equal to zero (see, e.g., the bottom panel of Figures~{\ref{SDO171_images}} \&~{\ref{SDO94_images}}). The numerical code then continues to loop over the number of chosen $x$ and $y$ values, cumulatively adding each subsequent histogram until the desired spatial size has been simulated. If the user has specified an optional base science image, then the program will automatically output a sample three-dimensional time series to convey the visual representation of the synthetic nanoflare fluctuations. To conserve the amount of disk space required, which becomes important when performing possibly millions of individual iterations of the code across all stipulated input parameters, the total amount of saved pixels is limited to $10^{6}$, which for a base science image of dimensions $350\times350$~pixels$^{2}$, provides a saved image sequence 50 frames long. These synthetic data products are important as one can play the simulated time series side-by-side with the observational data in order to visually ascertain the degree of similarity between the two image sequences.

\vspace{3mm}
In order to best characterize the resulting distribution, a number of statistical parameters are evaluated and saved to disk, including the median offset, Fisher and Pearson coefficients of skewness, a measurement of the kurtosis and the variance of the histogram, along with the width of the distribution at a variety of locations, including at half-maximum, quarter-maximum, eighth-maximum, etc. These measurements are contained within a single {\sc{idl}} save file, alongside a sample lightcurve for display purposes. Taking an observational sub-field as an example, a typical $350\times350$ pixel$^{2}$ field-of-view, each with $750$ consecutive frames (i.e., containing almost $1\times10^{8}$ individual pixels), takes on the order of $1.4\times10^{5}${\,}s (${\sim}$1.7{\,}days) on a 16-core 2.90~GHz Intel Xeon processor to fully generate, process, analyze and save the resulting outputs. The majority of this time is associated with the generation of the Poisson noise distributions, which to ensure a randomized sequence is produced for each new time series, requires $122{\,}500$ (i.e., $350\times350$) consecutive generations, each with $750$ individual values. 


\bibliographystyle{aasjournal.bst}

\end{document}